\documentclass[prd,showpacs,preprintnumbers,amsmath,amssymb,longbibliography, superscriptaddress, twocolumn, nofootinbib]{revtex4-1}
\usepackage{graphicx}
\usepackage{subcaption}
\usepackage{color}
\usepackage{hyperref}
\usepackage{cases}
\usepackage[utf8]{inputenc}

\usepackage{stackengine}
\stackMath
\usepackage{graphicx}

\newif\ifpreprint
\preprinttrue

\definecolor{airforceblue}{rgb}{0.36, 0.54, 0.66}
\definecolor{blue(ncs)}{rgb}{0.0, 0.53, 0.74}
\definecolor{caribbeangreen}{rgb}{0.0, 0.8, 0.6}

\def\be{\begin{equation}}
\def\ee{\end{equation}}
\newcommand\bea{\begin{align}}
\newcommand\eea{\end{align}}

\def\eqn#1{Equation~(\ref{#1})}

\def\eqns#1#2{Eqs.~(\ref{#1}) and~(\ref{#2})}

\def\sect#1{Section~{\ref{#1}}}

\newcounter{myExtFigureCounter}
\newcommand\getTikzFigure{{\includegraphics{"figures/loopCKMassive-figure\arabic{myExtFigureCounter}"}\stepcounter{myExtFigureCounter}}}

\begin{document}
\title{Loop-Level Double-Copy for Massive Quantum Particles}
\author{John Joseph M. Carrasco}
\affiliation{Department of Physics \& Astronomy, Northwestern University, Evanston, Illinois 60208, USA}
\affiliation{Institut de Physique Theorique, Universite Paris Saclay, CEA, CNRS, F-91191 Gif-sur-Yvette, France}
\author{Ingrid A. Vazquez-Holm}
\affiliation{Institut de Physique Theorique, Universite Paris Saclay, CEA, CNRS, F-91191 Gif-sur-Yvette, France}
\date{\today}
\begin{abstract}
We find  that scattering amplitudes in massive scalar QCD can manifest the duality between color and kinematics at loop-level. Specifically we construct the one-loop integrands for four-point scattering between two distinct massive scalars, and the five-point process encoding the first correction to massive scalar scattering with gluonic radiation. We find that factorization and the color-kinematics duality are sufficient principles to entirely bootstrap these calculations, allowing us to construct all contributions ultimately from the three-point tree-level amplitudes  which are themselves entirely constrained by symmetry. Double-copy construction immediately provides the associated predictions for massive scalars scattering in the so called ${\cal N}=0$ supergravity theory.  \end{abstract}

\maketitle


\section{Introduction}
\label{intro}

Traditional methods of calculating quantum gravitational scattering amplitudes using Feynman rules quickly run into difficulties.  There are two aspects to Feynman rule gravitational calculation that conspire to cause trouble. First, off-shell Feynman rules for gravitation are large and unwieldy, and have contact terms for every multiplicity of interaction.  Second, the scattering amplitudes involve a factorially growing number of local graphs -- either as multiplicity or loop level increases.  The former trouble can be alleviated by using unitarity methods -- the idea that one should only ever write down on-shell physical expressions when constructing predictions.  The latter can be mitigated by exploiting the double-copy structure to consider smaller building-blocks of physical predictions relevant to gravity, which are easier to calculate in gauge theories.  We will take advantage of both of these approaches to calculate $D$-dimensional scattering amplitudes in the $N=0$ supergravity theory through the integrand of one-loop five-point scattering between two distinct massive scalars with emitted gravitational radiation.  In the classical limit this will describe the gravitationally radiative correction to the scattering of scalar black holes. We will do this by bootstrapping $D$-dimensional predictions in massive scalar Quantum Chromodynamics (QCD).

Unitarity cuts~\cite{UnitarityMethod} allow for the integrands of incredibly complicated loop-level predictions to be verified systematically and invariantly with compact on-shell tree-level data.  Better yet, they allow for a principled fusing~\cite{fusing, BernMorgan} of tree-level data into multi-loop integrands. 
This is true independent of presence or absence of supersymmetry, independent of number and representation of colors, or flavors, and independent of number or type of massive particles.  Ref.~\cite{Bern:2007ct}  inverted the verification of generalized unitarity cuts~\cite{BCFUnitarity, Ossola:2006us, Forde:2007mi} towards a constructive approach that was ansatz free (cf.~\cite{Bern:2017ucb}), an approach called the {\em method of maximal cuts}.    This does not mitigate against the factorial explosion in graphs, but it means every stage of construction need only deal with a subset of graphs, and only their compact on-shell expressions. 

Double-copy structure is more subtle, but may ultimately prove just as useful for higher order corrections to phenomenological gravitational scattering.  We have known since the 1980s, from string theory limits, that color-stripped gauge theory semi-classical (tree-level) amplitudes encode gravitational amplitudes~\cite{KLT, Berends:1988zp}.  This realization is often informally referred to as the idea that ``Gravity is the square of Yang-Mills.''
One of the current authors, together with Bern and Johansson, discovered a local graphical story (BCJ) relating these predictions~\cite{Bern:2008qj}, called double-copy construction, one that generalizes straightforwardly to quantum (multi-loop) corrections~\cite{Bern:2010ue} at the integrand level.  

The key idea behind double-copy construction is that for gauge theory predictions (by definition containing both color and kinematic components), one can find representations where term by term kinematic weights obey the same algebraic relations as generic color weights.  Such kinematic weights are said to be color-dual. This makes  dynamics and charges interchangeable and indeed realizes  graviton (spin-2) scattering, in asymptotically flat space, as gluonic (spin-1) predictions whose   charges are the kinematics of gluons (spin-1).  This offers a profound calculational advantage even for gauge theories.  Because of the rigid algebraic rules locking the kinematic weights of each graph, only a small fraction of the graphs need to be dressed functionally, and all other local graphs inherit that dressing. Color-dual loop-level construction has largely occurred only in massless theories (or those where mass can be clearly associated with the dimensional reduction of a massless higher-dimensional theory). Here we make the novel demonstration that loop-level massive integrands in scalar-QCD admit color-dual descriptions.  This is a non-trivial step towards color-dual loop-level massive quark amplitudes.  

In four dimensions there are two physical states for every gluon, so one should expect four physical states in the double-copy (the outer product of gluonic states).  As (in four dimensions) gravitons also have two physical states, this means the natural double-copy in four-dimensions accounts for additional states beyond gravitons. In $D$ dimensions the double copy of gluonic states means tracking $(D-2)^2$ states (see discussions in  \cite{Bern:2019prr} and references therein).  Indeed, besides a graviton ($(D(D-3)/2)$ states), naive double-copy amplitudes generically will have a dilaton (1 state) as well as anti-symmetric tensors ($(D-2)(D-3)/2$ states) contributing to loop-predictions for massless external states.  In  gravitational double-copy amplitudes for massive external states,  even at tree level, one can generically expect the contributions of dilatons~\cite{Luna:2016hge}.  This state counting and attribution naturally fits in with the states that contribute to supergravity theories, which is why the naive double-copy of pure Yang-Mills is often called ${\cal N}=0$ supergravity.  This theory involves Einstein gravity coupled to a scalar field (known as the dilaton) as well as a two-form (often called the Kalb-Ramond field)---understandable as an axion in four dimensions.  Indeed the amplitudes presented in this paper double-copy to massive scalars coupled to the ${\cal N}=0$ supergravity theory.  Although it does not concern us for the purposes of this paper, it is worth pointing out that for pure gravities with less than half-maximal supersymmetry, including pure Einstein-Hilbert gravity, there are various strategies for projecting out such extra-gravitational  double-copy states~(see eg.~\cite{Johansson:2014zca, Luna:2017dtq, Bern:2019crd} and references therein).

Besides inviting a calculational virtuosity in gauge theories, there have been many motivations for studying quantum gravitational scattering amplitudes.  One recent driver of the field has been to understand the ultraviolet behavior of supergravity theories. This question is indeed responsible for the discovery of the duality between color and kinematics, as well as the associated double-copy construction, and has benefitted from it in return --- see e.g.~\cite{Bern:2010ue, SimplifyingBCJ,N46Sugra, N46Sugra2, Bern:2012cd,
  Bern:2012gh, Boels:2012sy, Bern:2013qca, Bern:2014lha, UVFiveLoops,
  Herrmann:2016qea, HerrmannTrnkaUVGrav}.  Even pure gravity manifests many unexpected cancellations beyond naive powercounting~\cite{Bern:2007xj}, suggesting that only a little  help in the ultraviolet may be sufficient for perturbative finiteness.    The advent of precision gravitational wave observation has provided a new urgency to discovering whether or not the simplicity in quantum gravitational gauge-invariant observable calculation can be applied to classical gravitational observables~(see e.g.~refs.~\cite{Saotome2012vy, Monteiro2014cda, Luna2015paa,Ridgway2015fdl, Luna2016due, White2016jzc,   Goldberger2016iau, Cardoso2016amd, Luna2016hge, Goldberger2017frp, Adamo2017nia, DeSmet2017rve, BahjatAbbas2017htu, 
  CarrilloGonzalez2017iyj, Goldberger2017ogt, Li2018qap,Ilderton:2018lsf,  ShenWorldLine,  Lee:2018gxc, Plefka:2018dpa, CheungPM, Berman:2018hwd, Gurses:2018ckx,  Adamo:2018mpq, Bahjat-Abbas:2018vgo, Luna:2018dpt,Kosower:2018adc,Farrow:2018yni, Bern:2019nnu, Bern:2019crd,Antonelli:2019ytb,CarrilloGonzalez:2019gof,Maybee:2019jus, PV:2019uuv, Huang:2019cja, Alawadhi:2019urr, Emond:2020lwi, Berman:2020xvs}).  Recent work has shown that multi-loop scattering amplitudes encode higher-order  corrections to classical observables. Indeed the highest order correction in the gravitational coupling, $G_N$, (often called post-Minkowskian [PM]) to conservative Black Hole binding energy (3 PM) was only made within the past couple of years and centered amplitudes insights (see ~\cite{Bern:2019nnu,Bern:2019crd} and references therein). This calculation required  the classical remnant of the two-loop four-point scattering amplitude between two massive scalars.  

Optimizing for the classical result, refs.~\cite{Bern:2019nnu,Bern:2019crd} exploit double-copy construction at tree-level, building the classically relevant gravitational integrand using unitarity methods from double-copied gravitational trees. The approach we present here is complementary. All unitarity construction occurs for the gauge theory only. The subsequent gravitational loop-level integrand arises from double-copy directly. In conjunction with an appropriate classicalization procedure, this approach offers  interesting possibilities. By lining up simultaneous classical Yang-Mills and Gravitational gauge-invariant observables where the duality between color and kinematics is manifest, one may hope to resolve ambiguities around applying double-copy directly in classical construction. We  leave associated extraction of classical predictions from the quantum integrands presented here to future consideration.

We will review color-kinematic construction in the adjoint and the fundamental in \sect{review}.  We will begin our bootstrap at tree-level in \sect{bootstrappingTrees} where we will find symmetry considerations alone completely fix the three-point amplitude, and higher multiplicity are entirely constrained by factorization and the color-kinematics duality.   Similarly we will only need to exploit these principles to construct our loop level results at four-point one-loop in \sect{fourPtOneLoop} and five-point one-loop in \sect{fivePtOneLoop}.  We  conclude and present next steps in \sect{conclusion}.
\section{Review: Color-Kinematics Duality from the Adjoint to the Fundamental}
\label{review}

From lining up predictions between Yang-Mills and gravity we find it useful to encode
predictions in mappings of cubic (tri-valent) graphs, a lesson that generalizes to the multi-loop level. Contact terms are encoded by allowing the cubic graphs to be dressed with inverse-propagators.  This also holds when dealing with massive matter in the fundamental~\cite{Naculich:2014naa, Johansson:2015oia, Luna:2017dtq, Plefka:2019wyg}. 
\begin{figure}
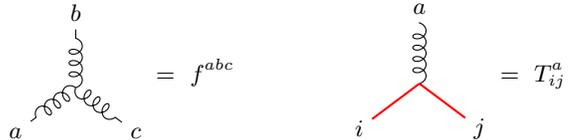

\begin{align*}
\begin{gathered}
\getTikzFigure
\end{gathered}
= ~  {f}^{abc} 
\hspace{1.5cm}
\begin{gathered}
\getTikzFigure
\end{gathered}
= ~T^a_{ij}
\end{align*}
\caption{Color-weights of cubic-vertices. The solid (red) lines represent massive scalars charged potentially under the fundamental representation of the group.}
\label{Fig:ColorCubic}
\end{figure}

\setcounter{myExtFigureCounter}{4}
The general form of a $m$-point $L$-loop gauge theory amplitude $\mathcal{A}^{(L)}_m$ in $D$ space-time dimensions with gluons and massive particles can be written as
\begin{equation}\label{Eq: General amplitude fraction}
\mathcal{A}^{(L)}_m = i^L g^{m-2+2L} \sum_{i \in \Gamma} \int \prod_{l=1}^L \frac{d^Dp_l}{(2 \pi)^D} \frac{1}{S_i} \frac{n_iC_i}{\prod_{\alpha_i}(p_{\alpha_i}^2 - m_{\alpha_i}^2)},
\end{equation}
where $g$ is the gauge coupling constant, and $m_{\alpha_i}^2$ is the on-shell mass of the particle with momentum $p_{\alpha_i}$. The sum runs over the complete set $\Gamma$ of $m$-point $L$-loop graphs with only cubic vertices, including all possible permutations of external legs. The integrations are over the independent loop momenta $p_l$, and each graph is dressed with a kinematic numerator $n_i$ unique to the graph topology, a color factor $C_i$ and the propagator structure of the graph.  The kinematic numerators $n_i$ are in general not gauge invariant objects. We take the convention of all external kinematics outgoing. At tree level \textit{color ordered amplitudes} can be constructed from  purely kinematic contributions (numerators over propagators) for graphs with the same external leg ordering\footnote{And therefore the same color factor.}.  Indeed these gauge-invariant expressions emerge naturally when expressing the color-weights in terms of a basis of color-factors in full tree-level amplitudes.  The kinematic coefficient of each basis color-weight will be a color-ordered amplitude.

The color factors are constructed from two types of terms: structure constants ${f}^{abc}$  
from purely gluonic vertices, and generators $T^a_{ij}$ from massive vertices, dressing the two types of cubic vertices (cf. Figure~\ref{Fig:ColorCubic}).  The structure constants are completely antisymmetric, and therefore the color-weights of adjoint vertices obey a flipping symmetry where exchanging the order of two of the legs in a vertex introduces a relative minus sign. As noted in \cite{Johansson:2015oia}, a convenient antisymmetry for fundamental generators similar to the structure constants can also be realized,
\begin{equation}
{f}^{abc} = -{f}^{bac} \hspace{1cm} {\rm and} \hspace{1cm} T^a_{ij} = -T^a_{ji}.
\end{equation}
From the structure of the group, the color factors obey Jacobi and analogous commutation relations associated with four-point subgraphs entirely in the adjoint, and in mixed adjoint-fundamental given by
\begin{equation}
\begin{aligned}
f^{dac}f^{cbe} - f^{dbc}f^{cae} &= f^{abc}f^{dce},\\
T^a_{ij}T^b_{jk} - T^b_{ij}T^a_{jk} &= f^{abc}T^c_{ik}.
\end{aligned}
\end{equation}

We will introduce an additional type of three-term  color identity, corresponding to a four-point sub-graph with all same-mass scalars. One can generalize the $SU(N)$ Fierz identity to optionally require
\begin{equation}
\label{potAdjointType}
T^a_{ij}T^a_{kl} = T_{ik}^b T_{jl}^b + T_{li}^c T_{jk}^c.
\end{equation}
This allows us the freedom to construct either adjoint scalar amplitudes or fundamental scalar amplitudes at our discretion\footnote{The kinematic weights of individual graphs will be applicable to either.}. We will refer to the set of these three-term identities as Jacobi-like relations.

\begin{figure}
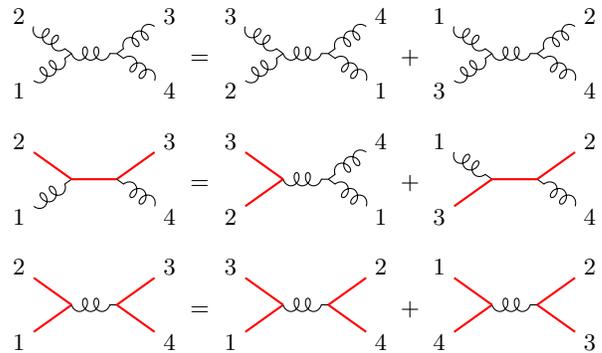

\begin{align*}
\begin{gathered}
\getTikzFigure
\end{gathered}
=
\begin{gathered}
\getTikzFigure
\end{gathered}
+
\begin{gathered}
\getTikzFigure
\end{gathered} 
\\
\begin{gathered}
\getTikzFigure
\end{gathered}
=
\begin{gathered}
\getTikzFigure
\end{gathered}
+
\begin{gathered}
\getTikzFigure
\end{gathered}
\\
\begin{aligned}
\begin{gathered}
\getTikzFigure
\end{gathered}
= 
\begin{gathered}
\getTikzFigure
\end{gathered}
+
\begin{gathered}
\getTikzFigure
\end{gathered}
\end{aligned}
\end{align*}
\caption{Pictorial representation of the relations between color weights and potentially kinematic weights of graphs.  The solid (red) lines represent same-mass scalar particles.}
\label{Fig: Graph representation of color kinematics}
\end{figure}

These relations are illustrated graphically in Figure~\ref{Fig: Graph representation of color kinematics} for gluons and massive scalar lines, and also represent the subsequent Jacobi-like and antisymmetry relations between color factors
\begin{equation}\label{Eq: Color jacobi and flipping}
c_i - c_j = c_k, \hspace{2cm} c_i \to -c_i.
\end{equation}
The relationship shown in Figure~\ref{Fig: Graph representation of color kinematics} applies to the four-point tree sub-graphs shown, but also to any set of three graphs that only differ by one internal leg with the three connectivities, where all other legs are held fixed.

In general, due to the cubic-assignment ambiguity of the four-point gluonic contact-term,  there is some degree of gauge freedom when choosing the kinematic and color numerator basis. The gauge invariance of an amplitude under different kinematic numerator choices is called \textit{generalized gauge freedom}. A fortuitous choice is a representation that obeys \textit{color kinematics duality}. This means choosing a representation where the kinematic numerators obey the same Jacobi-like relations and flipping symmetry as the color factors
\begin{equation}\label{Eq: color kinematic Jacobi and flipping}
\begin{aligned}
n_i - n_j = n_k \hspace{1cm} &\Leftrightarrow \hspace{1cm} c_i - c_j = c_k,\\
 n_i \to -n_i \hspace{1cm} & \Leftrightarrow \hspace{1cm} c_i \to -c_i
\end{aligned}
\end{equation}
in analogy to \eqn{Eq: Color jacobi and flipping}. These relations are known to hold up to all multiplicities at tree-level and conjectured to hold off-shell to all multiloop levels.  We will confirm this for one-loop at four and five-points.

Once we have found a representation of kinematic numerators that satisfy these relations, we can exchange the color factors in \eqn{Eq: General amplitude fraction} for another set of color-dual kinematic numerators 
\begin{multline}\label{Eq: Gravitational amplitude L-loop}
\mathcal{M}^{(L)}_m = i^{L+1} \Big( \frac{\kappa}{2} \Big)^{m-2+2L}\times\\
 \sum_{i \in \Gamma} \int \prod^L_{l=1} \frac{d^D p_l}{(2 \pi)^D} \frac{1}{S_i} \frac{n_i \tilde{n}_i}{\prod_{\alpha_i} (p_{\alpha_i}^2 - m_{\alpha_i}^2)},
\end{multline}
where the tilde on $\tilde{n}$ signifies that there can be two distinct gauge theories.  Indeed, only one of the numerator sets needs to manifest  the color kinematics duality. The expression obtained in \eqn{Eq: Gravitational amplitude L-loop}  is now a gravitational amplitude, and we call this relationship  the \textit{double copy construction} of gravitational amplitudes. There are many possible combinations of kinematic numerators from different gauge theories, with varying degrees of supersymmetry, which lead to amplitudes in different gravitational theories. However, in this paper we will focus on the double copy of scalar QCD theory to get  antisymmetric-tensor-dilaton-gravity coupled to massive scalars, recovering at tree-level amplitudes presented e.g. in ~\cite{Plefka:2019wyg}.
 
\section{Bootstrapping Tree Amplitudes}
\label{bootstrappingTrees}

The calculational setup for arriving at the tree level amplitudes in scalar QCD will be a bootstrapping method that takes advantage of factorization as well as the color-kinematics duality relations to fix kinematic ansatze for the graphs that contribute to a given amplitude.  Let us sketch in general how such a calculation proceeds, then consider specific calculation of relevant tree-level amplitudes needed to constrain via factorization the loop-level amplitudes we will eventually construct. 

\begin{figure}
\center
\begin{subfigure}{0.5\textwidth}
\includegraphics[scale=0.35]{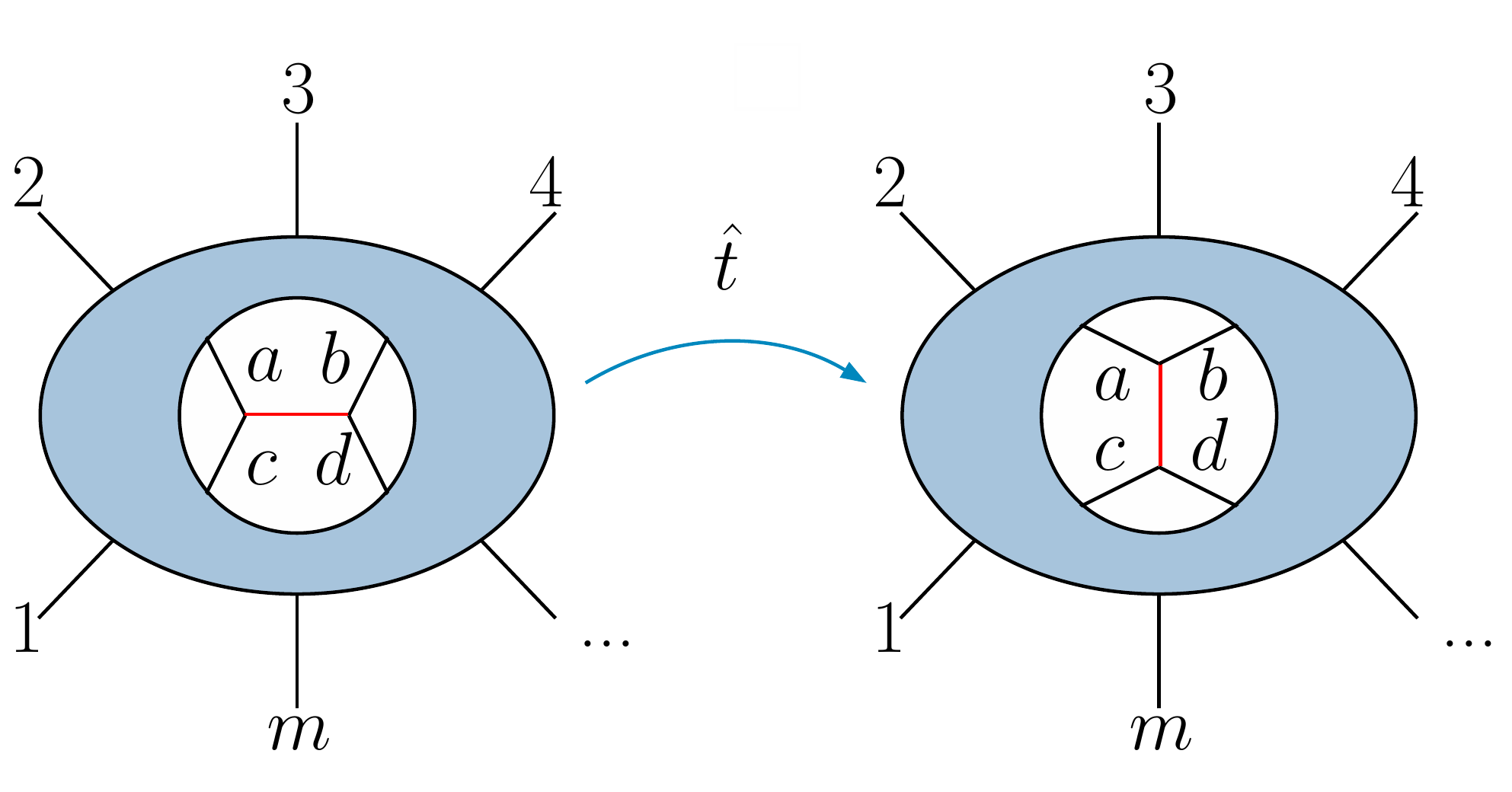}
\end{subfigure}
\begin{subfigure}{0.5\textwidth}
\includegraphics[scale=0.35]{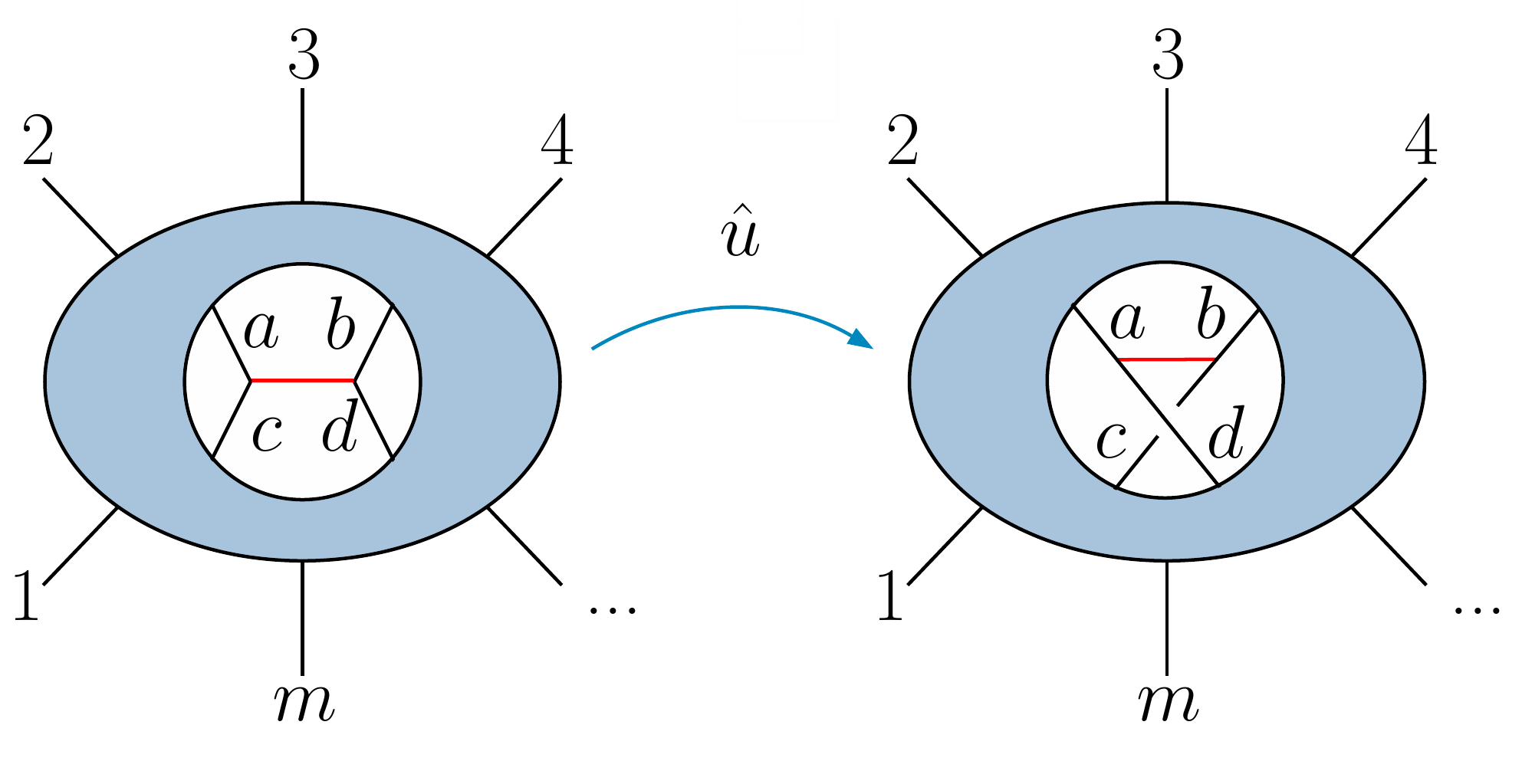}
\end{subfigure}
\caption{Graph manipulation operators that rearrange one internal edge's connectivity.  This is useful for specifying Jacobi-like relations, e.g. $c(g) = c(\hat{t}g) + c(\hat{u}g)$. }
\label{Fig:tuOperators}
\end{figure}

In general there will be some set of graphs $\Gamma$ contributing to a tree-level amplitude of a given multiplicity with particular external legs.  The graphs and color-kinematic relations between the numerators can be generated simultaneously.  As is standard in massless theories in the adjoint, one can introduce an operator that takes an edge of a graph, and returns a different graph which  is identical to the first graph except for the rearranged connectivity about the originally specified edge, as illustrated in Figure~\ref{Fig:tuOperators}.  Briefly summarized, a $\hat{t}$-operation takes an internal edge in a trivalent graph as a Mandelstam $s$-channel subgraph, and rearranges the connectivity to a Mandelstam $t$-channel subgraph, while holding all other legs fixed. Similarly, the $\hat{u}$-operator rearranges an $s$-channel to a $u$-channel subgraph.   So given an edge of one graph, one can understand the potential three-term identities it contributes to by simply considering for each edge:  $n(g) = n(\hat{t}\circ g) + n(\hat{u}\circ g)$.  By tracking any novel topologies introduced into a graph set by these operators, one can start with one tree-graph and generate all cubic graphs relevant to that amplitude simultaneously with all requisite color-dual kinematic constraints by operating on every edge of every graph until closure.

Once all the graphs of an amplitude are determined, the set of graphs can be arranged into a number of distinct topologies with different labelings. Each topology is assigned with a numerator function $n_i(k_1,...,k_m)$, which encodes all the kinematics of that specific topology, and is fully determined by the order of the $m$ external legs $(k_1,...,k_m)$. The generated color-dual  kinematic relations then become a system of equations between the numerator functions, which we can solve for in terms of the kinematic weights of  \textit{basis graph} topologies. All numerator functions can then be expressed as linear combinations of the basis graphs relabeled with different arguments. The kinematic weight of every basis graph topology is given an ansatz consisting of kinematic products of momenta and polarization vectors, depending on the amplitude in question.  For example, consider the purely gluonic amplitude.  For such an amplitude there is only one such basis graph at every multiplicity at tree-level, the so-called "half-ladder" or multi-peripheral graph.  All other topologies follow from $\hat{t}$ operations  (the color-order preserving whitehead move) away from the half-ladder.  The $\hat{u}$ operator on a half-ladder simply takes one to a half-ladder with permuted leg-label arguments.  This means for each multiplicity at tree-level for purely gluonic amplitudes, one need only supply an ansatz to a single topology.  We will see in various cases how this may change when we allow a combination of external massive scalars as well as glue.

To constrain the coefficients of the ansatze given to the basis graphs, we consider the relabeling symmetries -- or \textit{isomorphisms} --  of each topology. The set of all symmetries of all graphs, along with the kinematic duality relations which are not trivial after solving for the basis graphs, constrain the ansatz coefficients. One should note critically -- these are operations simply between functional numerators, not the factorial expanse of entire amplitudes nor even the relative exponential expanse of ordered amplitudes.  Any remaining parameters for this theory can be entirely constrained by factorization by considering all cuts of any single-propagator from all of the graph-topologies.  This color-kinematic bootstrap begins at the three-point tree-level. For scalar QCD amplitudes, these are entirely-fixed up to the coupling constant by mass dimension and anti-symmetry.  It should be noted, this means that for Yang-Mills amplitudes: factorization with color-dual kinematics encodes gauge-invariance.

\subsection{Three-point trees}
Let us consider first the purely-gluonic three-point amplitude. If it has a chance of being gauge invariant, every external polarization must appear in each term once and only once.   As all external legs are on-shell, conservation of momentum means any inner product between any two external momenta must vanish.  The Yang-Mills mass dimension allows us only three potential monomials in our basis:
\begin{multline}
n_3(k_1,k_2,k_3) = \alpha_1 (k_3 \cdot \epsilon_1)(\varepsilon_2 \cdot \varepsilon_3) \\+ \alpha_2 (k_3 \cdot \varepsilon_2)(\varepsilon_1 \cdot \varepsilon_3) 
+\alpha_3 (k_2 \cdot \varepsilon_3)(\varepsilon_1 \cdot \varepsilon_2) \,.
\end{multline}
Antisymmetry immediately constrains our ansatz via: $\alpha_2 \to - \alpha_3$, and $\alpha_1 \to \alpha_3$. We are left with only one free parameter $\alpha_3$ which can be taken to be the gluonic coupling $g$ appropriately scaled, fixed to canonical normalization by comparing with Feynman rules in a particular gauge (c.f. Feynman rules in e.g. refs.~\cite{Johansson:2015oia,  Plefka:2019wyg}), or set to unity, absorbing coupling constants in the definition of full amplitudes, and a phase in the definition of ordered amplitudes.   We choose the latter to minimize the complexity of kinematic numerator weights.

Next, to complete the three-point amplitudes for scalar QCD we need only consider the three-point amplitude with one external gluon and two external same-mass scalars.  Again we are not allowed any inner-products between external momenta, and we only have one polarization.  The mass-dimension means we can only write down a single inner-product, fortunately it itself is anti-symmetric via conservation of momenta: $k_1 \cdot \varepsilon_3 = -k_2 \cdot \varepsilon_3$ as $k_3\cdot \varepsilon_3=0$, yielding:
\begin{equation}
n_{3,2}(k_1^m, k_2^m,k_3) = \alpha_1 (k_1 \cdot \varepsilon_3) \,.
\end{equation}
Here, once again, we can take $\alpha_1$ to be the coupling constant suitably normalized, or as is our convention we will set $\alpha_1 =-1$, having pulled the coupling constant into the definition of the full amplitude.   We have written down all three-point amplitudes purely by considering mass-dimension and anti-symmetry up to normalization convention.  Everything else will follow through loop-level by simply considering factorization and the duality between color and kinematics. 

\subsection{Four-point trees}

\begin{figure}
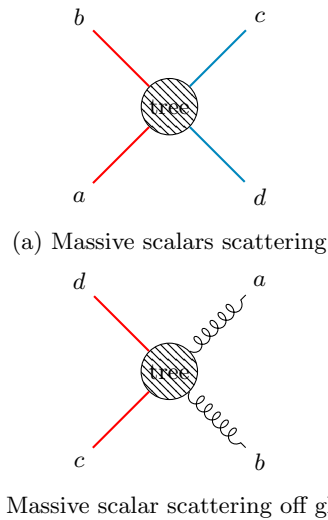

\begin{subfigure}{0.4\textwidth}
\getTikzFigure
\caption{Massive scalars scattering}
\end{subfigure}
\begin{subfigure}{0.4\textwidth}
\getTikzFigure
\caption{Massive scalar scattering off gluon}
\end{subfigure}
\caption{The four-point tree amplitudes in scalar QCD.}
\label{fig:fourPointTreeAmps}
\end{figure}

At four-point tree level in scalar QCD there are three distinct amplitudes.   The two distinct amplitudes that involve external masses:  one with one pair and one with two pairs of massive scalars respectively, are illustrated in Figure~\ref{fig:fourPointTreeAmps}.  The third amplitude, the purely gluonic amplitude, is not required for our one-loop construction so we do not report on it here, but it follows by the same bootstrap method we apply to these massive scalar amplitudes. 
\setcounter{myExtFigureCounter}{22}

\subsubsection{Two massive scalars}

The ordered four-point tree-amplitude for two pairs of massive scalars can be easily computed with a simple product of the three-point color-ordered Feynman rules~\cite{ Johansson:2015oia,  Plefka:2019wyg}.  The result, up to normalization and phase, is given by 
\begin{equation}\label{Eq: 4 massive scalars from feynman}
A_{4,2 \text{ (Feyn.)}}^{\text{tree}}(k_1^{m_1}
, k_2^{m_1}, k_3^{m_2}, k_4^{m_2}) = \frac{(k_1 - k_2)\cdot k_3}{2 (k_1+k_2)^2}.
\end{equation}
In this section we will introduce the bootstrapping framework amplitudes to reconstruct this amplitude -- overkill in this case, but as we will see this technique is quite generalizable and indeed takes us to loop-level.

\subsubsection{Bootstrapping the two pairs of massive scalars four-point amplitude}

Writing down the graphs that contribute to the two massive scalar pairs amplitude, we see that one graph \textit{topology}, or graph type, appears. The topology has a massless propagator, and for canonical external legs $(a, b, c, d)$ is given the kinematic numerator function 
\begin{align}\label{Eq: Numerator 4pt 2 massive scalars}
\begin{gathered}
\getTikzFigure
\end{gathered}
= n_{4,2}(a, b, c, d).
\end{align}
We note that this is the only graph that contributes to the amplitude, but to maintain bose-symmetry of the full-amplitude this kinematic weight must obey the same (anti-)symmetry properties as it's color weight with both massive scalars dressed in the adjoint.

We will now give an ansatz to the numerator function in \eqn{Eq: Numerator 4pt 2 massive scalars}. The graph has no external gluons, so the ansatz will consists solely of Lorentz products of momenta,
\begin{equation}\label{Eq: Ansatz 4pt 2 massive scalars}
n_{4,2}(a,b,c,d) = \alpha_1 t_{ab} + \alpha_2 t_{bb} + \alpha_3 t_{bc} + \alpha_4 t_{cc}, 
\end{equation}
where $t_{ij} \equiv (k_i \cdot k_j)$, and $\alpha_k$ are the free coefficients of the ansatz.

We determine some of the coefficients in \eqn{Eq: Ansatz 4pt 2 massive scalars} by using the isomorphisms of the graph in \eqn{Eq: Numerator 4pt 2 massive scalars}. The isomorphisms are expressed as relabelings of the numerator function
\begin{equation}\label{Eq: Symmetries 4pt 2 massive scalars}
\begin{gathered}
\begin{aligned}[c]
n_{4,2} &= -n_{4,2} (b, a, c, d),\\
n_{4,2} &= -n_{4,2} (a, b, d, c),\\
n_{4,2} &= \hspace{0.3cm}n_{4,2} (b, a, d, c),\\
n_{4,2} &=\hspace{0.3cm} n_{4,2} (c, d, a, b),
\end{aligned}
\qquad 
\begin{aligned}[c]
n_{4,2} &= -n_{4,2} (c, d, b, a),\\
n_{4,2} &= -n_{4,2} (d, c, a, b),\\
n_{4,2} &= \hspace{0.3cm}n_{4,2} (d, c, b, a),
\end{aligned}
\end{gathered}
\end{equation}
where we have suppressed the canonical arguments $(a, b, c, d)$ on the left-hand side of each equation. Writing out the symmetry relations in terms of the ansatz in \eqn{Eq: Ansatz 4pt 2 massive scalars} fixes three of the coefficients 
\begin{equation}
\alpha_4 = 0, \hspace{0.6cm} \alpha_1 = \alpha_2 = \frac{\alpha_3}{2},
\end{equation}
and the numerator function is determined up to an overall factor
\begin{equation}\label{Eq: 4 massive scalars ansatz after symmetries}
n_{4,2}(a,b,c,d) = \frac{\alpha_3}{2} \big( t_{ab} + t_{bb} + 2 t_{bc} \big).
\end{equation}

The numerator in \eqn{Eq: 4 massive scalars ansatz after symmetries} now obeys the symmetries of the graph for this amplitude.  The overall factor $\alpha_3$ is completely determined by considering the cut of the only propagator in the ordered amplitude,
\begin{equation}
A_{4,2}^{\text{tree}}(a^{m_1}, b^{m_1}, c^{m_2},d^{m_2} ) = \frac{n_{4,2}(a, b, c, d)}{(a+b)^2}.
\end{equation}
Satisfying factorization means that the sum over on-shell physical (cut) gluonic states of the three-point amplitudes is equal to $n_{4,2}$ evaluated under the cut condition:
\begin{multline}
\sum_{s \in {\rm states}} A_{3,2}(a^{m_1},b^{m_1},l^s) A_{3,2}(-l^{\overline{s}},c^{m_2},d^{m_2}) = \\
n_{4,2}(a,b,c,d) |_{(a+b)^2=0}^{\text{cut}}
\end{multline}
The sum runs over the polarization states of the $D$-dimensional polarization vectors and is given (e.g.~\cite{Bern:2019crd,Bern:2019prr,Kosmopoulos:2020pcd} and references therein) by the physical state projector
\begin{equation}\label{Eq: Physical state projector}
P^{\mu \nu}(p,q) = \sum_{\text{pols.}} \varepsilon^{\mu}(-p)\varepsilon^{\nu}(p) = \eta^{\mu \nu} - \frac{q^{\mu}p^{\nu}+p^{\mu}q^{\nu}}{q \cdot p},
\end{equation}
where $q$ is an arbitrary null reference momenta. Any such reference momenta must cancel out of any physical expression once the cut conditions and conservation of momenta have been imposed. It turns out for the cuts we consider in this paper, when each side is dressed with color-dual kinematic weights, the above general projector is equivalent to a much simpler gauge-dependent projector,
\begin{equation} \label{simpleProjector} 
\sum_{\text{pols.}} \varepsilon^{\mu}(-p)\varepsilon^{\nu}(p) \to \eta^{\mu \nu}\,.
\end{equation}  
This projector is a gauge-dependent choice, so if ever used in construction its appropriateness must be ensured~(c.f. \cite{Kosmopoulos:2020pcd}) or all gauge-compatibility must be verified via comparison with the above general projector (or equivalent constraints) as we have in this paper on a cut-by-cut basis. This completely fixes $\alpha_3 = -1$, matching via bootstrap exactly what is given by Feynman rules in \eqn{Eq: Ansatz 4pt 2 massive scalars}. The numerator function is therefore
\begin{equation}
n_{4,2}(a, b, c, d) = - \frac{1}{2} \big( t_{ab} + t_{bb} + 2 t_{bc} \big).
\end{equation}

The functional ordered amplitude for two massive scalar pairs at tree level follows from simply dividing the kinematic numerator by the single graph's massless propagator.

\subsubsection{One massive scalar}
\setcounter{myExtFigureCounter}{26}
The amplitude for one massive scalar pair scattering with a gluon represents a more lively application of bootstrapping approach, as it requires contributions from two distinct graph topologies whose kinematics are related by a color-dual identity. The relevant topologies are distinguished by whether the propagator is  massive , $M$, or massless, $\overline{M}$,
\begin{align}\label{Eq: Numerator 4pt 1 massive scalars}
\begin{gathered}
\getTikzFigure
\end{gathered}
\hspace{0.2cm} &= n_{4,1}^M(a,b,c,d)\, ,\\
\begin{gathered}
\getTikzFigure
\end{gathered}
&= n_{4,1}^{\overline{M}}(a, b, c, d).
\end{align}
It suffices to consider only one of the duality equations in order to identify a basis graph:
\begin{equation}
n_{4,1}^{\overline{M}}(a, b, c, d) =n_{4,1}^M(a, b, c, d) - n_{4,1}^M(b, a, c,d).
\end{equation} 
As $n_{4,1}^{\overline{M}}$ can be expressed entirely in terms of $n_{4,1}^M$, we identify $M$ as the basis graph and proceed to develope  a kinematic ansatz for its numerator weight, $n_{4,1}^M$.

As shown in \eqn{Eq: Numerator 4pt 1 massive scalars}, two of the external legs are gluons, so the ansatz will contain products of their polarization vectors. Each polarization vector appears once, and only once, in every term. We therefore apply the following ansatz
\begin{multline}
n_{4,1}^M(a, b, c, d) =\big(\alpha_1 t_{ab}  + \alpha_2 t_{bc} + \alpha_3 t_{cc} \big)t_{\varepsilon_a \varepsilon_b}+ \alpha_4 t_{b \varepsilon_a}t_{a \varepsilon_b} \\
+\alpha_5 t_{c \varepsilon_a}t_{a \varepsilon_b} +\alpha_6 t_{b \varepsilon_a}t_{c \varepsilon_b}+\alpha_7t_{c \varepsilon_a}t_{c \varepsilon_b}\, ,
\end{multline}
where $\alpha_i$ are the free coefficients of the ansatz, and $\varepsilon_p = \varepsilon(p)$ is the polarization vector of the particle with momenta $p$.

Similarly to the previous amplitude, we identify the isomorphisms of each of the contributing graphs,
\begin{equation}\label{Eq: Symmetries 4pt 1 massive scalars}
\begin{gathered}
\begin{aligned}[c]
n_{4,1}^{\overline{M}} &= -n_{4,1}^{\overline{M}}(b, a, c, d),\\
n_{4,1}^{\overline{M}} &= -n_{4,1}^{\overline{M}}(a,b, d, c),
\end{aligned}
\qquad 
\begin{aligned}[c]
n_{4,1}^{\overline{M}} &= n_{4,1}^{\overline{M}}(b, a, d, c),\\
n_{4,1}^M &= n_{4,1}^M(b, a, d, c),
\end{aligned}
\end{gathered}
\end{equation}
where we have again suppressed the canonical arguments $(a, b, c, d)$ on the left-hand side of each equation. Constraining the single ansatz in terms of these relations offers 
\begin{equation}
\alpha_5 = \alpha_7 -\alpha_6.
\end{equation}

To constrain the remaining coefficients of the numerator function one imposes factorization constraints as we did for the two-pairs of mass case considered earlier.  This yields,
\begin{equation}
\alpha_1=\alpha_3 =\alpha_4= 0, \hspace{0.4cm} \alpha_6 = \alpha_7= 2, \hspace{0.4cm} \alpha_2 = - 1\,.
\end{equation}
The basis numerator weight for the scattering of a massive scalar and gluon is then
\begin{equation}
\begin{aligned}
n_{4,1}^M(a, b, c, d) =t_{c  \varepsilon_b} (t_{b \varepsilon_a}+t_{c \varepsilon_a}) -\frac{1}{2}  t_{bc} t_{\varepsilon_a  \varepsilon_b} .
\end{aligned}
\end{equation}

The resulting ordered amplitude follows,
\begin{multline}
A^{\text{tree}}_{4,1}(a, b, c^{m}, d^{m}) = \frac{n_{4,1}^M(a, b, c,d)}{(b+c)^2-m^2}+\frac{n_{4,1}^{\overline{M}}(a, b, c,d)}{(a+b)^2}\\
= \frac{(t_{b\varepsilon_a}+t_{c\varepsilon_a})t_{c\varepsilon_b} - \frac{1}{2}t_{bc}t_{\varepsilon_a\varepsilon_b}}{(b+c)^2-m^2}+\\
\frac{t_{b\varepsilon_a}t_{c\varepsilon_b} - t_{c\varepsilon_a}t_{a\varepsilon_b}+\frac{1}{2}\left(t_{ac}-t_{bc} \right)t_{\varepsilon_a\varepsilon_b}
}{(a+b)^2}.
\end{multline}

The ordered amplitudes for a massive scalar and gluon scattering are known for explicit helicity configurations in $D=4$ spacetime dimensions. It is a useful exercise to compare our $D$-dimensional amplitude with appropriately chosen polarizations with the two independent four-dimensional amplitudes of ref.~\cite{Badger:2005zh}.

\subsection{Five-point trees}
\setcounter{myExtFigureCounter}{30}

\begin{figure}
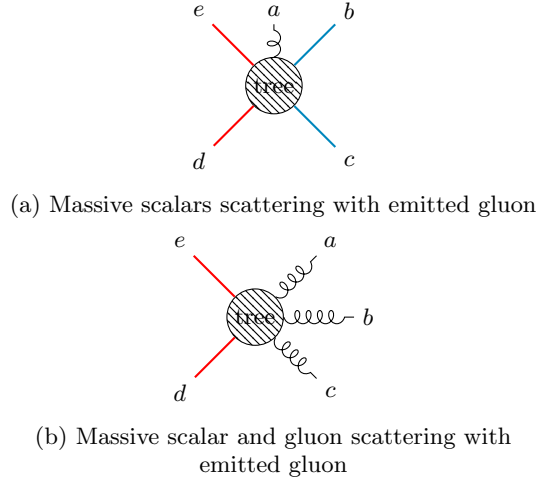

\begin{subfigure}{0.4\textwidth}
\getTikzFigure
\caption{Massive scalars scattering with emitted gluon}
\end{subfigure}
\begin{subfigure}{0.4\textwidth}
\getTikzFigure
\caption{Massive scalar and gluon scattering with emitted gluon}
\end{subfigure}
\caption{The five-point tree amplitudes in scalar QCD.}
\label{fig:fivePointTreeAmplitudes}
\end{figure}

At five-point tree level we again consider two distinct amplitudes with one and two massive scalar pairs, each now with an additional emitted gluon as shown in Figure~\ref{fig:fivePointTreeAmplitudes}. The amplitudes are bootstrapped using precisely the same procedure. The $D$-dimensional form of these tree level amplitudes we arrive at will allow a simple calculation of the  unitarity cuts required for loop level.

\subsubsection{Two massive scalars}

\begin{figure}
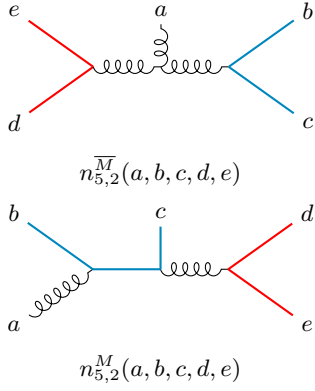

\begin{subfigure}{0.3\textwidth}
\getTikzFigure
\caption*{$n^{\overline{M}}_{5, 2}(a, b, c, d, e)$}
\end{subfigure}
~
\begin{subfigure}{0.3\textwidth}
\getTikzFigure
\caption*{$n_{5,2}^M(a, b, c, d, e)$}
\end{subfigure}
\caption{Graph topologies contributing to the amplitude for two massive scalars scattering with an emitted gluon.}
\label{Fig: Numerator 5pt 2 massive scalars}
\end{figure}

We consider first the five-point amplitude with two massive scalar pairs. The graph topologies for this amplitude are shown in Figure~\ref{Fig: Numerator 5pt 2 massive scalars}. We now distinguish between the topology with two massless propagators ($\overline{M}$), and the graph topology with one  of the  propagators massive ($M$).

To identify a basis graph for this amplitude we write down one of the duality equations, 
\begin{multline}
n_{5,2}^{\overline{M}}(a, b, c, d,e) = n_{5,2}^M(a, b, c, d,e)\\
-n_{5,2}^M(a, c,b, d,e),
\end{multline}
from which see the massless propagator graph's kinematic weight, $n_{5,2}^{\overline{M}}$, can be entirely expressed in terms of the $M$ graph's kinematics weight $n_{5,2}^M$. As such, we need only give the basis numerator $n_{5,2}^M$ an ansatz, which we will constrain via color-dual relations and factorization.  The rest of the amplitude follows directly.  The ansatz is given in terms of  Lorentz invariants of all momenta, including  the polarization vector of the external gluon, labelled $\varepsilon(a)$ in our canonical expression, which should appear once and only once in every term. We therefore begin with an ansatz of the form,
\begin{multline}\label{Eq: Ansatz 5pt 2 massive scalars}
n_{5,2}^M(a, b, c, d, e)=\\
t_{b\varepsilon_a} \left(\alpha_1 t_{ab}+\alpha_4 t_{ac}+\alpha_7 t_{ad} +\alpha_{10} t_{bc}+\alpha_{13} t_{cc}  \right. \\
\left. +\alpha_{16}  t_{cd}+\alpha_{19}  t_{dd} \right) \\
+t_{c\varepsilon_a} \left(\alpha_2 t_{ab}+\alpha_5 t_{ac}+\alpha_8 t_{ad}+\alpha_{11} t_{bc} +\alpha_{14} t_{cc} \right. \\
\left. +\alpha_{17}  t_{cd}+\alpha_{20} t_{dd}\right) \\
+ t_{d \varepsilon_a} \left(\alpha_3 t_{ab}
 +\alpha_6 t_{ac}+\alpha_9 t_{ad}+\alpha_{12} t_{bc}+\alpha_{15} t_{cc}\right. \\
\left.  +\alpha_{18}  t_{cd} +\alpha_{21} t_{dd}\right)\,.
\end{multline}

The isomorphisms of the graphs in Figure~\ref{Fig: Numerator 5pt 2 massive scalars} are 
\begin{equation}\label{Eq: Symmetries 5pt 2 massive scalars}
\begin{gathered}
\begin{aligned}[c]
n_{5,2}^M &= -n_{5,2}^M(a, b, c, e, d),\\
n_{5,2}^{\overline{M}} &= -n_{5,2}^{\overline{M}}(a, b, c, e, d),\\
n_{5,2}^{\overline{M}} &= -n_{5,2}^{\overline{M}}(a, c,b, d, e),\\
n_{5,2}^{\overline{M}} &= \hspace{0.3cm}n_{5,2}^{\overline{M}}(a, c,b, e, d),
\end{aligned}\,
\begin{aligned}[c]
n_{5,2}^{\overline{M}} &= -n_{5,2}^{\overline{M}}(a,d, e, b, c),\\
n_{5,2}^{\overline{M}} &= n_{5,2}^{\overline{M}}(a, d, e, c, b),\\
n_{5,2}^{\overline{M}} &= n_{5,2}^{\overline{M}}(a, e, d, b, c),\\
n_{5,2}^{\overline{M}} &= -n_{5,2}^{\overline{M}}(a, e, d, c, b)\,,
\end{aligned}
\end{gathered}
\end{equation}
where the canonical labeling $(a, b, c, d, e)$ has been suppressed on the left-hand sides of the equations. 

After imposing the symmetry constraints, eight coefficients remain to be determined.  The five-point ordered amplitude $A^{\text{tree}}_{5,2}(a, b, c, d, e)$ is given,
\begin{multline}
A^{\text{tree}}_{5,2}(a, b^{m_1}, c^{m_1}, d^{m_2}, e^{m_2}) = \frac{n_{5,2}^M (a, b, c, d, e)}{(d+e)^2\left((a+b)^2 - m_1^2\right)}\\ 
 + \frac{n_{5,2}^M(a, e, d, b, c)}{(b+c)^2\left((a+e)^2 -m_2^2\right)}+ \frac{n_{5,2}^{\overline{M}}(a, b, c, d, e)}{(b+c)^2(d+e)^2},
\end{multline}
where $m_1^2$ is the square mass of the scalar particle $b, c$, and $m_2^2$ is the square mass of the scalar particle $d, e$. Factorization involving one-particle cuts of both the massive propagator and the massles propagator fixes the remaining coefficients.

The basis numerator function then takes the simple form matching results in the literature~\cite{Luna2017dtq},
\begin{multline}
n_{5,2}^M(a, b, c, d, e) = \frac{1}{4}\left[ t_{ab}t_{c\varepsilon_a}+ 2 t_{ab} t_{d\varepsilon_a}+ \right.\\
\left.  \left(   t_{ab} +  2 t_{ac}  +  2 t_{bc} +  2 t_{cc}+4t_{cd} \right)t_{b\varepsilon_a} \right]\,,
\end{multline}
and the massless numerator function is given
\begin{multline}
n_{5,2}^{\overline{M}}(a, b, c, d, e) =\frac{1}{4} \left.[2 \left(t_{bc}+2 t_{cd}+t_{cc}\right) \left(t_{b\varepsilon_a}+t_{c\varepsilon_a}\right) \right.\\
\left. 
+t_{ab} \left(t_{b\varepsilon_a}+3 t_{c\varepsilon_a}+2 t_{d\varepsilon_a}\right) 
+t_{ac} \left(t_{b\varepsilon_a}+3 t_{c\varepsilon_a}-2 t_{d\varepsilon_a}\right) \right.
\\
\left. +4 t_{ad} t_{c\varepsilon_a}\right].
\end{multline}

\subsubsection{One massive scalar}

\begin{figure}
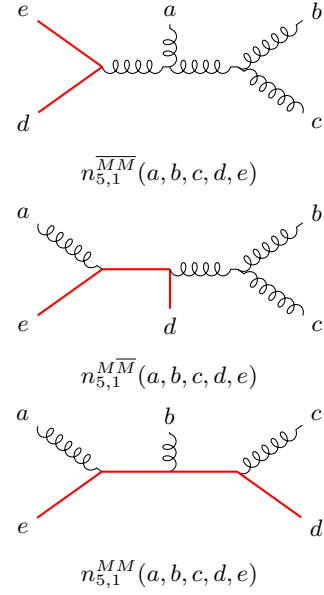

\begin{subfigure}{0.3\textwidth}
\getTikzFigure
\caption*{$n_{5, 1}^{\overline{M}\overline{M}}(a, b, c, d, e)$}
\end{subfigure}
~
\begin{subfigure}{0.3\textwidth}
\getTikzFigure
\caption*{$n_{5, 1}^{M\overline{M}}(a, b, c, d, e)$}
\end{subfigure}
~
\begin{subfigure}{0.3\textwidth}
\getTikzFigure
\caption*{$n_{5, 1}^{MM}(a, b, c, d, e)$}
\end{subfigure}
\caption{Graph topologies contributing to the amplitude for a massive scalar and gluon scattering, with an emitted gluon.}
\label{Fig:5pt1ms}
\end{figure}

The five-point amplitude with one massive scalar pair has three possible graph topologies, all shown in Figure~\ref{Fig:5pt1ms}. The topologies are distinguished by whether they have zero ($\overline{M}\overline{M}$), one ($M\overline{M}$), or two ($MM$) massive propagators. To determine the basis  graph under the kinematic algebra relations we need consider a subset of two kinematic color-dual relations, suppressing arguments $(a,b,c,d,e)$ on the LHS of the equations, 
\begin{align}
n_{5, 1}^{\overline{M}\overline{M}}&= 
n_{5, 1}^{M\overline{M}}(a, b, c, d, e) - n_{5, 1}^{M\overline{M}}(a, b, c, e, d),\\
 n_{5, 1}^{M\overline{M}}&=  n_{5, 1}^{MM}(a, b, c, d, e) - n_{5, 1}^{MM}(a, c, b, d, e). \nonumber
\end{align}
From the duality relations we see that ${MM}$ can be selected as the basis graph, and therefore proceed to give its numerator function a kinematic ansatz.

The amplitude has three external gluons, each of which should have a polarization vector in every term in the numerator ansatz.  We construct an 81 term ansatz for $n_5^{MM}$ from monomials of the forms $t_{i \varepsilon_j} t_{k \varepsilon_l} t_{m \varepsilon_n}$ and 
$t_{\varepsilon_i \varepsilon_j} t_{k \varepsilon_l} t_{mn}$.

The graphs in Figure~\ref{Fig:5pt1ms} have the following isomorphisms
\begin{align}
n_{5, 1}^{\overline{M}\overline{M}} &= - n_{5, 1}^{\overline{M}\overline{M}}(a, b, c, e, d), &
n_{5, 1}^{\overline{M}\overline{M}} &= - n_{5, 1}^{\overline{M}\overline{M}}(a, c, b, d, e), \nonumber \\
n_{5, 1}^{\overline{M}\overline{M}} &= \hspace{0.3cm} n_{5, 1}^{\overline{M}\overline{M}}(a, c, b, e, d),&
n_{5, 1}^{M\overline{M}} &= - n_{5, 1}^{M\overline{M}}(a, c, b, d, e), \nonumber \\
n_{5, 1}^{MM} &= - n_{5, 1}^{MM}(c, b, a, e, d)\,, &&
\end{align}
where the canonical labeling $(a, b, c, d, e)$ has been omitted on the left-hand side of the equations. Imposing the symmetries fixes 38 of the 81 coefficients of the ansatz, the remaining physical 40 parameters are fixed by factorization involving only  one-particle cuts.  It is amusing to note that three remaining unconstrained coefficients  never show up in a physical amplitude, representing residual generalized gauge freedom available to this five-point color-dual representation.  The resulting $D$-dimensional ordered amplitude is given by 
\begin{multline}
A^{\text{tree}}_{5,1}(a, b, c, d^{m}, e^{m}) = \frac{n_{5,1}^{MM}(a, b, c, d, e)}{\left((c+d)^2-m^2\right)\left((a+e)^2-m^2\right)}\\
+ \frac{n_{5,1}^{M\overline{M}}(a, b, c, d, e)}{\left((a+e)^2-m^2\right)(b+c)^2}+ \frac{n_{5,1}^{\overline{M}\overline{M}}(a, b, c, d, e)}{(b+c)^2(d+e)^2}\\
- \frac{n_{5,1}^{M\overline{M}}(c, b, a, e, d)}{\left((c+d)^2-m^2\right)(a+b)^2}
+ \frac{n_{5,1}^{\overline{M}\overline{M}}(c, b, a, d, e)}{(d+e)^2(a+b)^2}.
\end{multline}
This amplitude can be numerically compared with the 4D amplitudes given in \cite{Badger:2005zh}\footnote{Verifying all of \cite{Badger:2005zh}'s five-point three-glue and two-scalar amplitudes, with a corrected relative sign for the case where two of the external gluons are positive and one negative.}

\section{Bootstrapping One-loop Integrands}

\begin{figure}
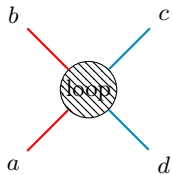

\getTikzFigure
\caption{The four-point loop amplitude for massive scalar scattering  in scalar QCD.}
\label{Figure:4 point 1 loop amplitude}
\end{figure}

\subsection{Loops: just like trees!}

Because the bootstrap approach -- applied at tree-level in the previous section -- is functional in terms of graph topologies, we can apply it directly towards constructing loop integrands.  At loop level  the role of tree-level one-particle factorization considerations is replaced by more general multi-leg unitarity cuts. At one-loop the numerator functions now take an additional argument compared to the same-multiplicity tree level case, namely the loop momentum $\ell$. The kinematic Jacobi-like relations are used to find a minimal set of basis graphs, which are in turn dressed with ansatze consistent with expected power counting. The ansatze are finally constrained using the remaining kinematic Jacobi-like relations, symmetries of the graph topologies, and a spanning set of generalized unitarity cuts.  

Recall that the full gauge amplitudes will be given by \eqn{Eq: General amplitude fraction}, and the full ${\cal N}=0$ supergravity amplitudes given by \eqn{Eq: Gravitational amplitude L-loop}.  The subsequent sections will be concerned with identifying the appropriate color-dual kinematic weights, $n(g)$, of each graph topology relevant to one-loop four-point, and one-loop five-point amplitudes respectively.

\subsection{Four-point One-Loop Construction}
\label{fourPtOneLoop}
\begin{figure*}
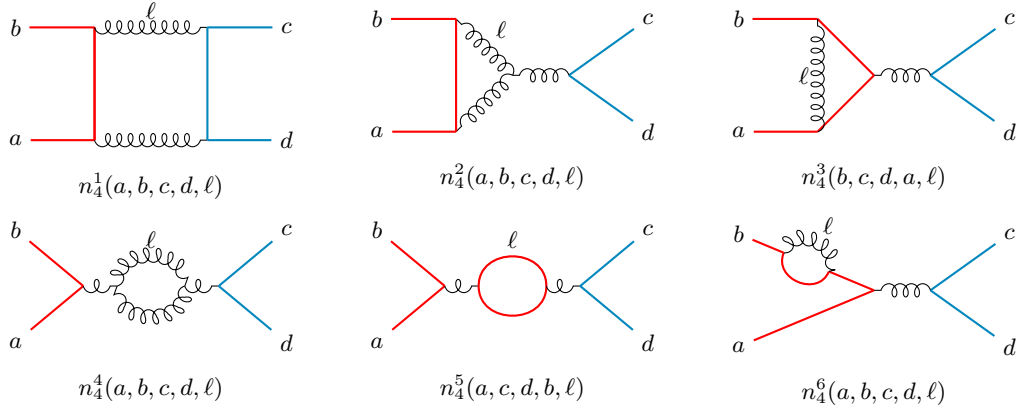

\begin{subfigure}{0.25\textwidth}
\getTikzFigure
\caption*{$n_4^1(a, b, c, d,\ell)$}
\end{subfigure}
~
\begin{subfigure}{0.25\textwidth}
\getTikzFigure
\caption*{$n_4^2(a, b, c, d, \ell)$}
\end{subfigure}
~
\begin{subfigure}{0.25\textwidth}
\getTikzFigure
\caption*{$n_4^3( b, c,d,a, \ell)$}
\end{subfigure}

\begin{subfigure}{0.25\textwidth}
\getTikzFigure
\caption*{$n_4^4(a, b, c, d, \ell)$}
\end{subfigure}
~
\begin{subfigure}{0.25\textwidth}
\getTikzFigure
\caption*{$n_4^5(a, c, d,b, \ell)$}
\end{subfigure}
~
\begin{subfigure}{0.25\textwidth}
\getTikzFigure
\caption*{$n_4^6(a, b, c, d, \ell)$}
\end{subfigure}
\caption{Graph topologies that contribute to $\mathcal{A}_4^{\text{1-loop}}$.}
\label{Fig: 1 loop 4 point graph topologies}
\end{figure*}

In this section we determine a color-dual representation of the four-point one-loop amplitude for two pairs of massive scalars. The possible graph topologies of this amplitude are shown in Figure~\ref{Fig: 1 loop 4 point graph topologies}, including the snail graph\footnote{We refer to graphs with a bubble on an external leg as snail graphs.} and excluding tadpoles. By inspecting the kinematic duality relations we find that we have some freedom in choosing the basis graphs for this amplitude.  We will take as basis graphs: $n_4^1(a, b, c, d, \ell)$ and $n_4^3(a, b, c, d, \ell)$. The remaining numerator functions can be written as linear combinations of these,
\begin{align}
n_4^2&\equiv n_4^1(a, b, c, d, c + d - \ell) - n_4^1(a, b, d, c, c + d - \ell), \nonumber \\
n_4^4 &\equiv n_4^1(a, b, c, d, c + d - \ell) + n_4^1(a, b, c, d, \ell) \nonumber\\
& - n_4^1(a, b, d, c, c + d - \ell) - n_4^1(a, b, d, c, \ell),\\
 n_4^5 &\equiv n_4^3(a, c, d, b, a - \ell) 
+ n_4^3(a, c, d, b, -b + \ell), \nonumber\\ 
n_4^6&\equiv  \left( n_4^1(a, b,c, d, \ell) -  d\leftrightarrow c \right) -n_4^3(a, c, d, b, \ell)  \nonumber\,,
\end{align}
where all kinematic numerator weights on the LHS of the equalities functionally take arguments:
$(a, b, c, d, \ell)$.

The basis graphs $n_4^1(a, b, c, d, \ell)$ and $n_4^3(a, b, c, d, \ell)$ are given ansatze, based on power counting and the known form of tree level amplitudes in the previous section. All external particles are scalars, and we expect to see a monomial of degree two in Lorentz invariants per term.  The number of independent monomials of the form $t_{ab} t_{cd}$ including inner products with loop momenta $\ell$ give 36 parameter ansatz for $n_4^1$.  The basis numerator $n_4^3$ is the same size, so the total number of free coefficients is therefore 72. As with the tree-level amplitudes we impose symmetries and color-kinematic constraints (see, e.g. Appendix~\ref{fourPointOneLoopJacobiAppendix}), which fix 43 of the coefficients. This leaves 29 coefficients to constrain via generalized unitarity cuts.

\begin{figure*}
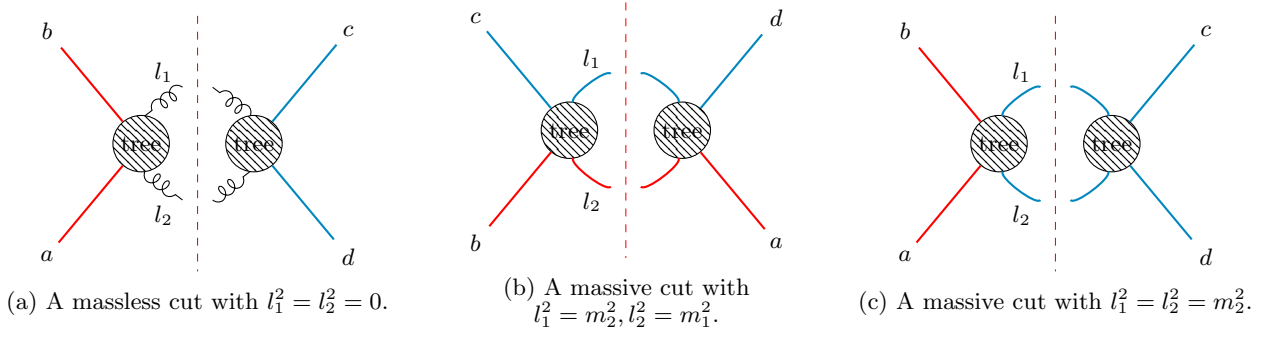

\begin{subfigure}{0.3\textwidth}
\getTikzFigure
\caption{A massless cut with $l_1^2 = l_2^2 = 0$.}
\label{Fig: 4 pt 1 loop massless cut}
\end{subfigure}
~
\begin{subfigure}{0.3\textwidth}
\getTikzFigure
\caption{A massive cut with $l_1^2 = m_2^2, l_2^2 = m_1^2$.}
\label{Fig: 4 pt 1 loop massive cut m1 m2}
\end{subfigure}
~
\begin{subfigure}{0.3\textwidth}
\getTikzFigure
\caption{A massive cut with $l_1^2 = l_2^2 = m_2^2$.}
\label{Fig: 4 pt 1 loop massive cut m1}
\end{subfigure}
\caption{Generalized unitarity cuts on the four-point one-loop amplitude.}
\label{Figure: cuts on four-point one loop}
\end{figure*}

The unitarity cuts are performed using the tree amplitudes generated in the previous section. Three unitarity cuts will be sufficient to determine the physical part of the numerators\footnote{The numerators contain some generalized gauge freedom that cannot contribute to any integrated expression.}, and they are illustrated in Figure~\ref{Figure: cuts on four-point one loop}. The first unitarity cut puts two massless internal legs on-shell, and the second and third cuts put two massive internal legs on-shell with \textit{distinct} and \textit{equal} masses, respectively.

The massless ordered cut in Figure~\ref{Fig: 4 pt 1 loop massless cut} gets contributions from  $n_4^1$, $n_4^2$ and $n_4^4$, and can be written as graphs dressed with numerator functions and corresponding propagator structures 
\begin{multline}
A_4^{\text{1-loop}}(a^{m_1}, b^{m_1}, c^{m_2}, d^{m_2})\Big|^{\text{cut}}_{\substack{l_1^2 \to 0\\ l_2^2 \to 0}} =\\
\frac{n_4^2(a, b, c, d, -a-b-l_1)}{((b+l_1)^2 - m_1^2)(a+b)^2} -\frac{n_4^2(d, c, a, b, a+b+l_1)}{((c-l_1)^2-m_2^2)(a+b)^2}\\
+ \frac{n_4^4(a, b, c, d, l_1)}{\left((a+b)^2\right)^2} 
+ \frac{n_4^1(a, b, c, d, l_1)}{((b+l_1)^2 - m_1^2)((c-l_1)^2 -m_2^2)} ,
\end{multline}
where $m_1^2$ is the squared mass of the scalar $a, b$ and $m_2^2$ is the squared mass of the other scalar $c, d$. To determine coefficients, but also to verify the validity of our numerators so far, this cut should be equal to the product of two four-point trees with one massive scalar. We determined these tree amplitudes  in the previous section, allowing us to easily work out,
\begin{multline}
A_4^{\text{1-loop}}(a^{m_1}, b^{m_1}, c^{m_2}, d^{m_2})\Big|_{\substack{l_1^2 \to 0\\l_2^2 \to 0 }}^{\text{cut}} =\\
 \sum_{s_1, s_2} A^{\text{tree}}_{4,1} (a^{m_1}, b^{m_1}, l_1^{s_1}, l_2^{s_2}) A^{\text{tree}}_{4,1} (-l_2^{\bar{s}_2}, -l_1^{\bar{s}_1}, c^{m_2}, d^{m_2}) ,
\end{multline}
where the sum is over the possible states of the gluonic cut legs $l_1, l_2$. The sum runs over the polarization states of the $D$-dimensional polarization vectors just as we did when considering factorization at tree-level, so we can employ the same physical projectors (e.g. \eqns{Eq: Physical state projector}{simpleProjector}).

To further determine the coefficients of the amplitude we perform the massive cut in Figure~\ref{Fig: 4 pt 1 loop massive cut m1 m2}. This cut only gets a contribution from the box graph, which is now dressed in the following way
\begin{equation}
A_4^{\text{1-loop}}(a^{m_1}, b^{m_1}, c^{m_2}, d^{m_2})\Big|^{\text{cut}}_{\substack{l_1^2 \to m_1^2\\ l_2^2 \to m_2^2}} = \frac{n_4^1 (a, b, c, d, l_1-b)}{(b-l_1)^2(a+l_1)^2}.
\end{equation}
On the tree side of this cut, we get a product of the four-point trees with two massive scalars from the previous section,
\begin{multline}
A_4^{\text{1-loop}}(a^{m_1}, b^{m_1}, c^{m_2}, d^{m_2})\Big|^{\text{cut}}_{\substack{l_1^2 \to m_1^2\\ l_2^2 \to m_2^2}}  =  \\
A^{\text{tree}}_{4, 2} (a^{m_1}, l_1^{m_1}, l_2^{m_2}, d^{m_2}) A^{\text{tree}}_{4, 2} (-l_2^{m_2}, -l_1^{m_1}, b^{m_1}, c^{m_2}). 
\end{multline}
The cut legs are now massive scalars with only one allowed state \footnote{We are free to consider cuts where we fix the mass of the particle crossing the cut.}, so the state sum is trivial and the cut fixes five coefficients. 

After the massive and massless cut we are left with 13 undetermined coefficients. Some of these are fixed on the third and final cut shown in Figure~\ref{Fig: 4 pt 1 loop massive cut m1}. This unitarity cut puts two massive internal lines on shell, now of the same type. There are two possibilities, $l_1^2 = l_2^2 = m_1^2$ and $l_1^2 = l_2^2 = m_2^2$, and we need only consider the latter, e.g. ${l_1^2, l_2^2 \to m_2^2}$. Once this set of cuts is satisfied it fixes an additional eight coefficients. Because of the symmetry of the amplitude and the individual graphs, the corresponding unitarity cuts with $l_1^2 = l_2^2 = m_1^2$ is also satisfied but puts no further constraints on the ansatz. Finally the amplitude numerators are determined up to five free coefficients, which represent generalized gauge freedom that cannot contribute to any integrated expression.

It is worth discussing this last cut in a little more detail.   What do we mean by an ordered four-point tree when all external scalar masses are the same?  This is the only distinction for one-loop four-point that occurs depending on when we allow the color version of \eqn{potAdjointType} to be satisfied.  If we are to allow this, our ordered amplitudes can be thought of as {\em adjoint-type}, so e.g. we can have two channels contribute to an ordered color-stripped amplitude:
\begin{equation}
A^{\rm tree}_{4,{\rm adj}}(1^{m_1},2^{m_1},3^{m_1},4^{m_1}) = \frac{n_s}{s}+\frac{n_t}{t}\,.
\end{equation}
where 
\begin{align}
n_s &=   n_{4,2}(1^{m_1},2^{m_1},3^{m_1},4^{m_1})\, ,\\
n_t &=  n_{4,2}(4^{m_1},1^{m_1},2^{m_1},3^{m_1}) , \nonumber
\end{align}
and $s,t$ are Mandelstam invariants:
 \begin{align}
s&=(k_1+k_2)^2\, , \\
t&=(k_2+k_3)^2\,. \nonumber
\end{align}
  
   If on the other hand, we demand that external scalars labeled 1 and 3 are particles distinct from anti-particle legs labeled 2 and 4, all still with the same mass, we have two distinct ordered color-stripped amplitudes with one channel each: 
\begin{align}
A^{\rm tree}_{4,2}(1^{m_1},\bar{2}^{m_1},3^{m_1},\bar{4}^{m_1}) &= \frac{n_s}{s} \\
A^{\rm tree}_{4,2}(1^{m_1},\bar{4}^{m_1},3^{m_1},\bar{2}^{m_1}) &= \frac{n_t}{t}\,. \nonumber
\end{align}
We can refer to these types of ordered trees as {\em fundamental-type}.  As is perhaps not so surprising, the same kinematic weights written functionally can be used for both theories, the amplitudes differ simply in which graphs one allows to contribute.   Demanding the kinematic analog of \eqn{potAdjointType} for functional numerators:
\begin{multline}
  n_{4,2}(1^{m_1},2^{m_1},3^{m_1},4^{m_1}) =
  n_{4,2}(3^{m_1},1^{m_1},4^{m_1},2^{m_1}) + \\
  n_{4,2}(4^{m_1},1^{m_1},2^{m_1},3^{m_1}) \, ,
\end{multline}
is simply an additional constraint that can be imposed on the kinematic dressings.  If we could not satisfy these types of conditions, we would be unable to satisfy adjoint-type cuts, but it turns out both for four-point one-loop and five-point one-loop there is no problem establishing such constraints.  As a result our same kinematic weights satisfy both adjoint-type and fundamental-type cuts, and indeed one can use adjoint-type cuts to constrain all parameters in our-bootstrap.   The pattern we see here of adjoint-type ordered amplitudes being sums over individually gauge-invariant fundamental-type ordered amplitudes persists to higher multiplicity. 

\subsection{Five-point One-Loop Const ruction}

\label{fivePtOneLoop}

In this section we will consider the five-point one-loop amplitude with two pairs of massive scalars. Compared to the construction of the four-point loop amplitude in the previous section, at five-point the number of graph topologies is much larger and the external gluon contributes with a polarization vector that makes the form of the numerators more complicated.

The five-point one-loop amplitude has 33 possible graph topologies, shown in Figures~\ref{fivePointGraphsPartI}-\ref{fivePointGraphsPartII}, again including snails but not tadpoles. The two massive scalars are illustrated by a blue and red line representing the different masses, and we impose that the numerators be symmetric under the exchange of these lines. Note that some of the graphs with purely massive loops belong to two independent topologies that depend on the mass of the internal loop. For example, $n_5^{24}$ and $n_5^{25}$ are two distinct topologies. This plays a role in determining the Jacobi-like relationship between the graphs. Solving for these Jacobi-like relations we find that a possible set of basis graphs is $n_5^1$, $n_5^2$, $n_5^8$, $n_5^{14}$, $n_5^{15}$ and $n_5^{33}$. Each basis graph is given a non-zero kinematic ansatz.

The ansatze for the kinematic numerators must contain the polarization vector of the external gluon, $\varepsilon_a$, and have the correct power counting. We therefore propose the following form
\begin{equation}
n^{\lambda}_5(a, b, c, d, e, \ell) = \sum_{p=1}^{312} \alpha^{\lambda}_{p} ~(k_i \cdot k_j)(k_l \cdot k_m) (k_n \cdot \varepsilon_a), 
\end{equation}
where $\alpha^{\lambda}_{p}$ are the free coefficients, and $k_a$ are the momenta of external particles or loop momenta. Each ansatz has 312 terms, so the total number of coefficients is 1872 for all six ansatze.


\begin{figure*}
\begin{subfigure}[b]{0.25\textwidth}
\getTikzFigure
\caption*{$n_5^1(a, e, b, c, d,\ell)$}
\end{subfigure}
~
\begin{subfigure}[b]{0.25\textwidth}
\getTikzFigure
\caption*{$n_5^2(a, b, c, d,e, \ell)$}
\end{subfigure}
~
\begin{subfigure}[b]{0.25\textwidth}
\getTikzFigure
\caption*{$n_5^3(a, b, c, d, e, \ell)$}
\end{subfigure}

\begin{subfigure}[b]{0.25\textwidth}
\getTikzFigure
\caption*{$n_5^4(a, b, c, d, e,\ell)$}
\end{subfigure}
~
\begin{subfigure}[b]{0.25\textwidth}
\getTikzFigure
\caption*{$n_5^5(a, d, b, c, e, \ell)$}
\end{subfigure}
~
\begin{subfigure}[b]{0.25\textwidth}
\getTikzFigure
\caption*{$n_5^6(a, e, b, c, d,\ell)$}
\end{subfigure}

\begin{subfigure}[b]{0.25\textwidth}
\getTikzFigure
\caption*{$n_5^7(a, b, c, d,e, \ell)$}
\end{subfigure}
~
\begin{subfigure}[b]{0.25\textwidth}
\getTikzFigure
\caption*{$n_5^8(a, b, c, d, e,\ell)$}
\end{subfigure}
~
\begin{subfigure}[b]{0.25\textwidth}
\getTikzFigure
\caption*{$n_5^9(a, b, c, d,e, \ell)$}
\end{subfigure}

\begin{subfigure}[b]{0.25\textwidth}
\getTikzFigure
\caption*{$n_5^{10}(a, b, c, d,e, \ell)$}
\end{subfigure}
~
\begin{subfigure}[b]{0.25\textwidth}
\getTikzFigure
\caption*{$n_5^{11}(a, d, b, c, d, \ell)$}
\end{subfigure}
~
\begin{subfigure}[b]{0.25\textwidth}
\getTikzFigure
\caption*{$n_5^{12}(a, b, c, d,e, \ell)$}
\end{subfigure}

\begin{subfigure}[b]{0.25\textwidth}
\getTikzFigure
\caption*{$n_5^{13}(a, b, c, d, e,\ell)$}
\end{subfigure}
~
\begin{subfigure}[b]{0.25\textwidth}
\getTikzFigure
\caption*{$n_5^{14}(a, e,b, c, d,\ell)$}
\end{subfigure}
~
\begin{subfigure}[b]{0.25\textwidth}
\getTikzFigure
\caption*{$n_5^{15}(a, b, c, d,e, \ell)$}
\end{subfigure}

\begin{subfigure}[b]{0.25\textwidth}
\getTikzFigure
\caption*{$n_5^{16}(a, b, c, d,e,  \ell)$}
\end{subfigure}
~
\begin{subfigure}[b]{0.25\textwidth}
\getTikzFigure
\caption*{$n_5^{17}(a, b, c, d, e,\ell)$}
\end{subfigure}
~
\begin{subfigure}[b]{0.25\textwidth}
\getTikzFigure
\caption*{$n_5^{18}(a, b, c, d,e, \ell)$}
\end{subfigure}
\caption{One-loop five-point graph topologies (excluding tadpoles) related by Jacobi and commutation relations (1-18).}
\label{fivePointGraphsPartI}
\end{figure*}

\begin{figure*}
\begin{subfigure}[b]{0.25\textwidth}
\getTikzFigure
\caption*{$n_5^{19}(a, d, b, c, e, \ell)$}
\end{subfigure}
~
\begin{subfigure}[b]{0.25\textwidth}
\getTikzFigure
\label{ex:snail}
\caption*{$n_5^{20}(a, b, c, d,e, \ell)$}
\end{subfigure}
~
\begin{subfigure}[b]{0.25\textwidth}
\getTikzFigure
\caption*{$n_5^{21}(a, b, c,d, e,  \ell)$}
\end{subfigure}

\begin{subfigure}[b]{0.25\textwidth}
\getTikzFigure
\caption*{$n_5^{22}(a, b, c, d,e, \ell)$}
\end{subfigure}
~
\begin{subfigure}[b]{0.25\textwidth}
\getTikzFigure
\caption*{$n_5^{23}(a, b, c, d, e, \ell)$}
\end{subfigure}
~
\begin{subfigure}[b]{0.25\textwidth}
\getTikzFigure
\caption*{$n_5^{24}(a, b, c, d, e, \ell)$}
\end{subfigure}

\begin{subfigure}[b]{0.25\textwidth}
\getTikzFigure
\caption*{$n_5^{25}(a, b, c, d, e, \ell)$}
\end{subfigure}
~
\begin{subfigure}[b]{0.25\textwidth}
\getTikzFigure
\caption*{$n_5^{26}(a, b, c, d, e, \ell)$}
\end{subfigure}
~
\begin{subfigure}[b]{0.25\textwidth}
\getTikzFigure
\caption*{$n_5^{27}(a, b, c, d, e, \ell)$}
\end{subfigure}

\begin{subfigure}[b]{0.25\textwidth}
\getTikzFigure
\caption*{$n_5^{28}(a, b, c, d, e, \ell)$}
\end{subfigure}
~
\begin{subfigure}[b]{0.25\textwidth}
\getTikzFigure
\caption*{$n_5^{29}(a, e, b, c, d, \ell)$}
\end{subfigure}
~
\begin{subfigure}[b]{0.25\textwidth}
\getTikzFigure
\caption*{$n_5^{30}(a, b, c, d,e, \ell)$}
\end{subfigure}

\begin{subfigure}[b]{0.25\textwidth}
\getTikzFigure
\caption*{$n_5^{31}(a, b, c, d,e, \ell)$}
\end{subfigure}
~
\begin{subfigure}[b]{0.25\textwidth}
\getTikzFigure
\caption*{$n_5^{32}(a, e, b, c, d,  \ell)$}
\end{subfigure}
~
\begin{subfigure}[b]{0.25\textwidth}
\getTikzFigure
\caption*{$n_5^{33}(a, b, c, d, e, \ell)$}
\end{subfigure}
\caption{One-loop five-point graph topologies (excluding tadpoles) related by Jacobi and commutation relations (19-33).}
\label{fivePointGraphsPartII}
\end{figure*}

Imposing the symmetries of all graphs, along with the remaining relationships from the Jacobi-like relations, constrains 1190 of the coefficients. After imposing the symmetries, there are then 682 coefficients left to determine using generalized unitarity cuts.

The ordered unitarity cuts required to constrain the five-point amplitude are graphically depicted in Figure~\ref{Figure: cuts on five-point one loop}.   One new feature at five-point one-loop is that ability to have two-particle cuts (bubble cuts) where one of the cut legs is massive and one is massless (c.f Figure~\ref{Fig: 5 pt 1 loop one massive one massless cut}). Here are all the cuts that must be performed expressed in terms of the tree amplitudes that contribute:
\begin{widetext}
\begin{align}
&A^{\text{1-loop}}_5(a, b^{m_1}, c^{m_1}, d^{m_2}, e^{m_2}) \Big|_{ \substack{l_1^2 \to 0\\ l_2^2 \to 0 }}^{\text{cut}} =  \sum_{s_1,s_2}A^{\text{tree}}_{4, 1}(d^{m_2}, e^{m_2},l_1^{s_1}, l_2^{s_2}) A^{\text{tree}}_{5, 1}(-l_2^{\overline{s_2}}, -l_1^{\overline{s_1}}, b^{m_1}, a, c^{m_1})\,, \\
&A^{\text{1-loop}}_5(a, b^{m_1}, c^{m_1}, d^{m_2}, e^{m_2}) \Big|^{\text{cut}}_{ \substack{l_1^2 \to m_{1}^2 \\ l_2^2 \to 0} }  =  \sum_{s_2} A^{\text{tree}}_{4, 1}(a, b^{m_1},l_1^{m_1}, l_2^{s_2}) A^{\text{tree}}_{5, 2}(-l_2^{\overline{s_2}}, -l_1^{m_1}, c^{m_1}, d^{m_2}, e^{m_2})\,, \\
&A^{\text{1-loop}}_5(a, b^{m_1}, c^{m_1}, d^{m_2}, e^{m_2}) \Big|^{\text{cut}}_{\substack{l_1^2 \to m_2^2\\l_2^2 \to m_1^2 }} = A^{\text{tree}}_{4, 2}(c^{m_1}, d^{m_2}, l_1^{m_2}, l_2^{m_1}) A^{\text{tree}}_{5, 2}(-l_2^{m_1}, -l_1^{m_2}, e^{m_2}, a, b^{m_1})\,, \\
&A^{\text{1-loop}}_5(a, b^{m_1}, c^{m_1}, d^{m_2}, e^{m_2}) \Big|^{\text{cut}}_{ \substack{l_1^2 \to m_{2}^2 \\ l_2^2 \to m_2^2} } =  A^{\text{tree}}_{4, 2}(d^{m_2}, e^{m_2}, l_1^{m_2}, l_2^{m_2}) A^{\text{tree}}_{5, 2}(-l_2^{m_2}, -l_1^{m_2}, a, b^{m_1}, c^{m_1}) \, , \\
&A^{\text{1-loop}}_5(a, b^{m_1}, c^{m_1}, d^{m_2}, e^{m_2}) \Big|^{\text{cut}}_{ \substack{l_1^2 \to m_{1}^2 \\ l_2^2 \to m_1^2} } =  A^{\text{tree}}_{4, 2}(d^{m_2}, e^{m_2}, l_1^{m_1}, l_2^{m_1}) A^{\text{tree}}_{5, 2}(-l_2^{m_1}, -l_1^{m_1}, a, b^{m_1}, c^{m_1}) \,.
\end{align} 
\end{widetext}
After these cuts all but 123 of our original parameters are constrained.  This is sufficient to ensure that all physical cuts are satisfied. The resulting integrand with all color-dual consistent gauge freedom is included in an ancillary file with the arXiv version of this paper. 

Note that the above cuts involve ordered partial amplitudes.  It may seem surprising that we require no additional partial-amplitude orderings under the same kinematic constraints. After all at five-point tree-level, for a single pair of massive scalars,  there are two independent ordered partial-amplitudes under amplitude relations.  We witness here one of the benefits of using functional kinematic graph dressings tightly constrained by algebraic relations.  In general one need only consider a smaller set of cuts (or factorization channels as at tree-level) to entirely constrain an integrand.  Indeed for the maximally supersymmetric gauge theory at three-loop four-points only one maximal cut of one graph is required to constrain the entire color-dual integrand~\cite{Carrasco:2011hw}.  If one is concerned that additional ordered cuts may be required one can always examine the full color-dressed cut. Equivalently, for each set of kinematic cut conditions, one can verify that no additional orderings are required by simply considering the associated double-copy gravitational cut and noting the vanishing of any coefficients of remaining ansatz parameters.  In any case we verify our final integrand on a set of physical spanning cuts as we discuss in the next section.

\begin{figure*}
\begin{subfigure}[t]{0.3\textwidth}
\getTikzFigure
\caption{A massless cut with $l_1^2 = l_2^2 = 0$.}
\label{Fig: 5 pt 1 loop massless cut}
\end{subfigure}
~
\begin{subfigure}[t]{0.3\textwidth}
\getTikzFigure
\caption{A massless-massive cut with $l_1^2 = m_1^2$ and $ l_2^2 = 0$.}
\label{Fig: 5 pt 1 loop one massive one massless cut}
\end{subfigure}

\begin{subfigure}[t]{0.3\textwidth}
\getTikzFigure
\caption{A massive cut with $l_1^2 = m_2^2$ and $ l_2^2 = m_1^2$.}
\label{Fig: 5 pt 1 loop massive cut m1 m2}
\end{subfigure}
~
\begin{subfigure}[t]{0.3\textwidth}
\getTikzFigure
\caption{A massive cut with $l_1^2 = l_2^2 = m_2^2$.}
\label{Fig: 5 pt 1 loop massive cut m2x2}
\end{subfigure}
~
\begin{subfigure}[t]{0.3\textwidth}
\getTikzFigure
\caption{A massive cut with $l_1^2 = l_2^2 = m_1^2$.}
\label{Fig: 5 pt 1 loop massive cut m1x1}
\end{subfigure}
\caption{Generalized unitarity cuts on the five-point one-loop amplitude.}
\label{Figure: cuts on five-point one loop}
\end{figure*}

\subsection{Verification}

For both four-point one-loop and five-point one-loop, we consider the potential reduction to an integral basis by verifying the above integrands on all bubble, box, and triangle cuts with distinct color-orders under the following restrictions:
\begin{itemize}
\item  Every bubble cut (two-particle cut) must involve a physical tree amplitude (multiplicity four or higher). 
\item  Every triangle cut (three-particle cut) must involve the loop momenta flowing through each tree.
\item Every box cut (four-particle cut) must involve the loop-momenta flowing through each tree. 
\end{itemize}
This set of restrictions precludes snail diagrams (e.g. diagrams 18, 19, 20, 28, 29, 30, 31, 32, and 33 of Figures~\ref{fivePointGraphsPartI} and \ref{fivePointGraphsPartII}) as well as tadpole diagrams from contributing to the cuts. As practical applications of the gravitational double-copy of these amplitudes will involve the $\hbar\to0$ limit, we expect this to be completely sufficient~\cite{Bern:2019crd}.   We verify on this spanning set of cuts that no remaining parameters contribute to physical cuts, and can be taken as pure gauge choice relative to any set of related physical observables.

We should emphasize that  there are important questions around how to appropriately consider snail and tadpole contributions to the UV.  We refer the interested reader to ref.~\cite{BernMorgan} for analysis handling similar issues in the context of unitarity based approaches.

\section{Conclusion}
\label{conclusion}

In this paper, we frame a constructive bootstrap solely in terms of factorization and color-dual representations for $D$-dimensional massive scalar QCD amplitudes, leading to  integrands at one loop in massive scalar QCD.  Specifically we found the first loop correction to scattering between two different massive scalar fields, as well as the first loop correction to such scattering with an emitted gluon.  Along the way we recalculated a number of tree-level amplitudes in this theory, starting with the three-point amplitudes constrained entirely by mass-dimension and symmetry.  By exploiting the duality between color and kinematics -- known to hold at tree-level, but still conjectural at loop level, we found tremendously simplified representations -- requiring a small number of basis graph topologies whose functional kinematic weights encode the entire amplitude.   We confirmed the validity of this approach at tree-level by comparing with results in the literature, and at loop level on a spanning set of physical cuts.  Besides the computational simplicity in the gauge theory, and further evidence for the conjecture of loop-level color-dual representations, a critical advantage of our current approach is that once the gauge amplitudes have been constructed in this color-dual form, building the corresponding gravity integrand in ${\cal N}=0$ supergravity is trivial: one simply exchanges the color-weights with kinematics weights graph by graph.

The gauge theory representations we constructed in this paper have a number of intriguing properties. Notably these representations land naturally on a gauge that plays well with summing Yang-Mills states over a simplified projector~(cf. \eqn{simpleProjector} relative to \eqn{Eq: Physical state projector}) which aids in unitarity-based constructions.  Additionally color-dual representations propagate cut information to graphs with ambiguous or ill-defined cuts like massive snail or tadpole graphs. While these do not contribute to physical cuts relevant to classical limits, such graphs will be relevant to quantum UV behavior and mass renormalization.  This implies a generalized gauge choice.  It will be intriguing to apply the approach presented here to massive quark amplitudes in QCD with care to reproduce standard regularization schemes.

There is significant current interest in massive scattering amplitudes in pure Einstein-Hilbert gravity due to relevance to precision gravitational wave physics.  A natural next step will be to take our ${\cal N}=0$ supergravity double-copy results and project out any unwanted massless states.  More non-trivial will be to generalize this type of multiloop scattering amplitude methods to massive particles with arbitrary spin.  While ambitious, important clarifying progress is already being made  both in field theory and scattering communities (see, e.g., \cite{Holstein:2008sx,Vaidya:2014kza,Guevara:2017csg,Bini:2017xzy,Vines:2018gqi,Guevara:2018wpp,Chung:2018kqs,Guevara:2019fsj,Johansson:2019dnu,Chung:2019duq,Damgaard:2019lfh,Bautista:2019evw,Aoude:2020onz,Bern:2020buy}, and references therein).  We expect color-dual methods to play an increasingly vital role in this exploration.

In summary, the combination of unitarity methods and the duality between color and kinematics is incredibly powerful.  While originally discovered in the quest of understanding the UV behavior of maximally supersymmetric gravity theory in four-dimensions, the duality between color and kinematics is both dimension-agnostic as well as entirely independent of the presence or absence of supersymmetry.   It is gratifying to discover that massive scattering amplitudes at loop-level can be compatible with these principles, and we look forward to continuing to extend our visibility into loop-level prediction now in increasingly more phenomenological theories. 

\section{Acknowledgements}
We thank Zvi Bern, Henrik Johansson, David Kosower, Donal O'Connell, Ben Page, Nic Pavao, Sebastian P\"{o}gel, Laurentiu Rodina, Radu Roiban, Aslan Seifi, Bogdan Stoica,  and Suna Zekio\u{g}lu for many useful and interesting related discussions. This project has received funding from the European Union’s Horizon 2020 research and innovation programme under the Marie Skłodowska-Curie grant agreement No. 764850 (SAGEX).  JJMC was supported for much of this work by the European Research Council under ERC-STG-639729, Strategic Predictions for Quantum Field Theories. IVH is supported by the SAGEX ITN. 

\appendix

\section*{Appendix}
\subsection{Four-point one-loop: Kinematic Jacobi-like Relations}
\label{fourPointOneLoopJacobiAppendix}

Here we tabulate the Jacobi relations satisfied by four-point one-loop, two pairs of massive external scalars.  Five-point one-loop is included in a machine readable ancillary file.
\begin{widetext}
\begin{equation}
\begin{aligned}
n_4^2(a,b,c,d,\ell)&=n_4^6(a,b,c,d,-a-b-\ell)+n_4^3(a,c,d,b,-a-b-\ell),\\
n_4^2(a,b,c,d,\ell)&=-n_4^2(a,b,c,d,c+d-\ell)+n_4^4(a,b,c,d,c+d-\ell),\\
n_4^2(a,b,c,d,\ell)&=-n_4^6(b,a,c,d,\ell)+n_4^3(a,c,d,b,-\ell),\\
n_4^2(a,b,c,d,\ell)&=n_4^1(a,b,c,d,c+d-\ell)-n_4^1(a,b,d,c,c+d-\ell),\\
n_4^6(a,b,c,d,\ell)&=n_4^2(a,b,c,d,c+d-\ell)-n_4^3(a,c,d,b,\ell),\\
n_4^3(a,c,d,b,\ell)&=n_4^2(a,b,c,d,-\ell)+n_4^6(b,a,c,d,-\ell),\\
n_4^3(a,c,d,b,\ell)&=-n_4^3(a,c,d,b,a-b-\ell)+n_4^5(a,c,d,b,b+\ell),\\
n_4^3(a,c,d,b,\ell)&=n_4^2(a,b,c,d,c+d-\ell)-n_4^6(a,b,c,d,\ell),\\
n_4^4(a,b,c,d,\ell)&=n_4^2(c,d,a,b,-c-d+\ell)-n_4^2(d,c,a,b,-c-d+\ell),\\
n_4^4(a,b,c,d,\ell)&=n_4^2(a,b,c,d,c+d-\ell)+n_4^2(a,b,c,d,\ell),\\
n_4^1(a,b,c,d,\ell)&=n_4^2(c,d,a,b,-\ell)+n_4^1(a,b,d,c,c+d-\ell),\\
n_4^1(a,b,c,d,\ell)&=n_4^2(a,b,c,d,c+d-\ell)+n_4^1(b,a,c,d,c+d-\ell),\\
n_4^5(a,c,d,b,\ell)&=n_4^3(a,c,d,b,-b-c-d-\ell)+n_4^3(a,c,d,b,a+c+d+\ell)
\end{aligned}
\end{equation}
\end{widetext}

\bibliography{loopCKMassive}

\begin{thebibliography}{87}%
\makeatletter
\providecommand \@ifxundefined [1]{%
 \@ifx{#1\undefined}
}%
\providecommand \@ifnum [1]{%
 \ifnum #1\expandafter \@firstoftwo
 \else \expandafter \@secondoftwo
 \fi
}%
\providecommand \@ifx [1]{%
 \ifx #1\expandafter \@firstoftwo
 \else \expandafter \@secondoftwo
 \fi
}%
\providecommand \natexlab [1]{#1}%
\providecommand \enquote  [1]{``#1''}%
\providecommand \bibnamefont  [1]{#1}%
\providecommand \bibfnamefont [1]{#1}%
\providecommand \citenamefont [1]{#1}%
\providecommand \href@noop [0]{\@secondoftwo}%
\providecommand \href [0]{\begingroup \@sanitize@url \@href}%
\providecommand \@href[1]{\@@startlink{#1}\@@href}%
\providecommand \@@href[1]{\endgroup#1\@@endlink}%
\providecommand \@sanitize@url [0]{\catcode `\\12\catcode `\$12\catcode
  `\&12\catcode `\#12\catcode `\^12\catcode `\_12\catcode `\%12\relax}%
\providecommand \@@startlink[1]{}%
\providecommand \@@endlink[0]{}%
\providecommand \url  [0]{\begingroup\@sanitize@url \@url }%
\providecommand \@url [1]{\endgroup\@href {#1}{\urlprefix }}%
\providecommand \urlprefix  [0]{URL }%
\providecommand \Eprint [0]{\href }%
\providecommand \doibase [0]{http://dx.doi.org/}%
\providecommand \selectlanguage [0]{\@gobble}%
\providecommand \bibinfo  [0]{\@secondoftwo}%
\providecommand \bibfield  [0]{\@secondoftwo}%
\providecommand \translation [1]{[#1]}%
\providecommand \BibitemOpen [0]{}%
\providecommand \bibitemStop [0]{}%
\providecommand \bibitemNoStop [0]{.\EOS\space}%
\providecommand \EOS [0]{\spacefactor3000\relax}%
\providecommand \BibitemShut  [1]{\csname bibitem#1\endcsname}%
\let\auto@bib@innerbib\@empty
\bibitem [{\citenamefont {Bern}\ \emph {et~al.}(1994)\citenamefont {Bern},
  \citenamefont {Dixon}, \citenamefont {Dunbar},\ and\ \citenamefont
  {Kosower}}]{UnitarityMethod}%
  \BibitemOpen
  \bibfield  {author} {\bibinfo {author} {\bibfnamefont {Zvi}\ \bibnamefont
  {Bern}}, \bibinfo {author} {\bibfnamefont {Lance~J.}\ \bibnamefont {Dixon}},
  \bibinfo {author} {\bibfnamefont {David~C.}\ \bibnamefont {Dunbar}}, \ and\
  \bibinfo {author} {\bibfnamefont {David~A.}\ \bibnamefont {Kosower}},\
  }\bibfield  {title} {\enquote {\bibinfo {title} {{One loop $n$-point gauge
  theory amplitudes, unitarity and collinear limits}},}\ }\href {\doibase
  10.1016/0550-3213(94)90179-1} {\bibfield  {journal} {\bibinfo  {journal}
  {Nucl. Phys.}\ }\textbf {\bibinfo {volume} {B425}},\ \bibinfo {pages}
  {217--260} (\bibinfo {year} {1994})},\ \Eprint
  {http://arxiv.org/abs/hep-ph/9403226} {arXiv:hep-ph/9403226 [hep-ph]}
  \BibitemShut {NoStop}%
\bibitem [{\citenamefont {Bern}\ \emph {et~al.}(1995)\citenamefont {Bern},
  \citenamefont {Dixon}, \citenamefont {Dunbar},\ and\ \citenamefont
  {Kosower}}]{fusing}%
  \BibitemOpen
  \bibfield  {author} {\bibinfo {author} {\bibfnamefont {Zvi}\ \bibnamefont
  {Bern}}, \bibinfo {author} {\bibfnamefont {Lance~J.}\ \bibnamefont {Dixon}},
  \bibinfo {author} {\bibfnamefont {David~C.}\ \bibnamefont {Dunbar}}, \ and\
  \bibinfo {author} {\bibfnamefont {David~A.}\ \bibnamefont {Kosower}},\
  }\bibfield  {title} {\enquote {\bibinfo {title} {{Fusing gauge theory tree
  amplitudes into loop amplitudes}},}\ }\href {\doibase
  10.1016/0550-3213(94)00488-Z} {\bibfield  {journal} {\bibinfo  {journal}
  {Nucl. Phys.}\ }\textbf {\bibinfo {volume} {B435}},\ \bibinfo {pages}
  {59--101} (\bibinfo {year} {1995})},\ \Eprint
  {http://arxiv.org/abs/hep-ph/9409265} {arXiv:hep-ph/9409265 [hep-ph]}
  \BibitemShut {NoStop}%
\bibitem [{\citenamefont {Bern}\ and\ \citenamefont
  {Morgan}(1996)}]{BernMorgan}%
  \BibitemOpen
  \bibfield  {author} {\bibinfo {author} {\bibfnamefont {Z.}~\bibnamefont
  {Bern}}\ and\ \bibinfo {author} {\bibfnamefont {A.~G.}\ \bibnamefont
  {Morgan}},\ }\bibfield  {title} {\enquote {\bibinfo {title} {{Massive loop
  amplitudes from unitarity}},}\ }\href {\doibase 10.1016/0550-3213(96)00078-8}
  {\bibfield  {journal} {\bibinfo  {journal} {Nucl. Phys.}\ }\textbf {\bibinfo
  {volume} {B467}},\ \bibinfo {pages} {479--509} (\bibinfo {year} {1996})},\
  \Eprint {http://arxiv.org/abs/hep-ph/9511336} {arXiv:hep-ph/9511336 [hep-ph]}
  \BibitemShut {NoStop}%
\bibitem [{\citenamefont {Bern}\ \emph {et~al.}(2007)\citenamefont {Bern},
  \citenamefont {Carrasco}, \citenamefont {Johansson},\ and\ \citenamefont
  {Kosower}}]{Bern:2007ct}%
  \BibitemOpen
  \bibfield  {author} {\bibinfo {author} {\bibfnamefont {Z.}~\bibnamefont
  {Bern}}, \bibinfo {author} {\bibfnamefont {J.~J.~M.}\ \bibnamefont
  {Carrasco}}, \bibinfo {author} {\bibfnamefont {Henrik}\ \bibnamefont
  {Johansson}}, \ and\ \bibinfo {author} {\bibfnamefont {D.~A.}\ \bibnamefont
  {Kosower}},\ }\bibfield  {title} {\enquote {\bibinfo {title} {{Maximally
  supersymmetric planar Yang-Mills amplitudes at five loops}},}\ }\href
  {\doibase 10.1103/PhysRevD.76.125020} {\bibfield  {journal} {\bibinfo
  {journal} {Phys. Rev.}\ }\textbf {\bibinfo {volume} {D76}},\ \bibinfo {pages}
  {125020} (\bibinfo {year} {2007})},\ \Eprint {http://arxiv.org/abs/0705.1864}
  {arXiv:0705.1864 [hep-th]} \BibitemShut {NoStop}%
\bibitem [{\citenamefont {Britto}\ \emph {et~al.}(2005)\citenamefont {Britto},
  \citenamefont {Cachazo},\ and\ \citenamefont {Feng}}]{BCFUnitarity}%
  \BibitemOpen
  \bibfield  {author} {\bibinfo {author} {\bibfnamefont {Ruth}\ \bibnamefont
  {Britto}}, \bibinfo {author} {\bibfnamefont {Freddy}\ \bibnamefont
  {Cachazo}}, \ and\ \bibinfo {author} {\bibfnamefont {Bo}~\bibnamefont
  {Feng}},\ }\bibfield  {title} {\enquote {\bibinfo {title} {{Generalized
  unitarity and one-loop amplitudes in ${\cal N}=4$ super-Yang-Mills}},}\
  }\href {\doibase 10.1016/j.nuclphysb.2005.07.014} {\bibfield  {journal}
  {\bibinfo  {journal} {Nucl. Phys.}\ }\textbf {\bibinfo {volume} {B725}},\
  \bibinfo {pages} {275--305} (\bibinfo {year} {2005})},\ \Eprint
  {http://arxiv.org/abs/hep-th/0412103} {arXiv:hep-th/0412103 [hep-th]}
  \BibitemShut {NoStop}%
\bibitem [{\citenamefont {Ossola}\ \emph {et~al.}(2007)\citenamefont {Ossola},
  \citenamefont {Papadopoulos},\ and\ \citenamefont {Pittau}}]{Ossola:2006us}%
  \BibitemOpen
  \bibfield  {author} {\bibinfo {author} {\bibfnamefont {Giovanni}\
  \bibnamefont {Ossola}}, \bibinfo {author} {\bibfnamefont {Costas~G.}\
  \bibnamefont {Papadopoulos}}, \ and\ \bibinfo {author} {\bibfnamefont
  {Roberto}\ \bibnamefont {Pittau}},\ }\bibfield  {title} {\enquote {\bibinfo
  {title} {{Reducing full one-loop amplitudes to scalar integrals at the
  integrand level}},}\ }\href {\doibase 10.1016/j.nuclphysb.2006.11.012}
  {\bibfield  {journal} {\bibinfo  {journal} {Nucl. Phys. B}\ }\textbf
  {\bibinfo {volume} {763}},\ \bibinfo {pages} {147--169} (\bibinfo {year}
  {2007})},\ \Eprint {http://arxiv.org/abs/hep-ph/0609007}
  {arXiv:hep-ph/0609007} \BibitemShut {NoStop}%
\bibitem [{\citenamefont {Forde}(2007)}]{Forde:2007mi}%
  \BibitemOpen
  \bibfield  {author} {\bibinfo {author} {\bibfnamefont {Darren}\ \bibnamefont
  {Forde}},\ }\bibfield  {title} {\enquote {\bibinfo {title} {{Direct
  extraction of one-loop integral coefficients}},}\ }\href {\doibase
  10.1103/PhysRevD.75.125019} {\bibfield  {journal} {\bibinfo  {journal} {Phys.
  Rev. D}\ }\textbf {\bibinfo {volume} {75}},\ \bibinfo {pages} {125019}
  (\bibinfo {year} {2007})},\ \Eprint {http://arxiv.org/abs/0704.1835}
  {arXiv:0704.1835 [hep-ph]} \BibitemShut {NoStop}%
\bibitem [{\citenamefont {Bern}\ \emph {et~al.}(2017)\citenamefont {Bern},
  \citenamefont {Carrasco}, \citenamefont {Chen}, \citenamefont {Johansson},
  \citenamefont {Roiban},\ and\ \citenamefont {Zeng}}]{Bern:2017ucb}%
  \BibitemOpen
  \bibfield  {author} {\bibinfo {author} {\bibfnamefont {Zvi}\ \bibnamefont
  {Bern}}, \bibinfo {author} {\bibfnamefont {John Joseph~M.}\ \bibnamefont
  {Carrasco}}, \bibinfo {author} {\bibfnamefont {Wei-Ming}\ \bibnamefont
  {Chen}}, \bibinfo {author} {\bibfnamefont {Henrik}\ \bibnamefont
  {Johansson}}, \bibinfo {author} {\bibfnamefont {Radu}\ \bibnamefont
  {Roiban}}, \ and\ \bibinfo {author} {\bibfnamefont {Mao}\ \bibnamefont
  {Zeng}},\ }\bibfield  {title} {\enquote {\bibinfo {title} {{Five-loop
  four-point integrand of $N=8$ supergravity as a generalized double copy}},}\
  }\href {\doibase 10.1103/PhysRevD.96.126012} {\bibfield  {journal} {\bibinfo
  {journal} {Phys. Rev.}\ }\textbf {\bibinfo {volume} {D96}},\ \bibinfo {pages}
  {126012} (\bibinfo {year} {2017})},\ \Eprint
  {http://arxiv.org/abs/1708.06807} {arXiv:1708.06807 [hep-th]} \BibitemShut
  {NoStop}%
\bibitem [{\citenamefont {Kawai}\ \emph {et~al.}(1986)\citenamefont {Kawai},
  \citenamefont {Lewellen},\ and\ \citenamefont {Tye}}]{KLT}%
  \BibitemOpen
  \bibfield  {author} {\bibinfo {author} {\bibfnamefont {H.}~\bibnamefont
  {Kawai}}, \bibinfo {author} {\bibfnamefont {D.~C.}\ \bibnamefont {Lewellen}},
  \ and\ \bibinfo {author} {\bibfnamefont {S.~H.~H.}\ \bibnamefont {Tye}},\
  }\bibfield  {title} {\enquote {\bibinfo {title} {{A relation between tree
  amplitudes of closed and open strings}},}\ }\href {\doibase
  10.1016/0550-3213(86)90362-7} {\bibfield  {journal} {\bibinfo  {journal}
  {Nucl. Phys.}\ }\textbf {\bibinfo {volume} {B269}},\ \bibinfo {pages} {1--23}
  (\bibinfo {year} {1986})}\BibitemShut {NoStop}%
\bibitem [{\citenamefont {Berends}\ \emph {et~al.}(1988)\citenamefont
  {Berends}, \citenamefont {Giele},\ and\ \citenamefont
  {Kuijf}}]{Berends:1988zp}%
  \BibitemOpen
  \bibfield  {author} {\bibinfo {author} {\bibfnamefont {Frits~A.}\
  \bibnamefont {Berends}}, \bibinfo {author} {\bibfnamefont {W.T.}\
  \bibnamefont {Giele}}, \ and\ \bibinfo {author} {\bibfnamefont
  {H.}~\bibnamefont {Kuijf}},\ }\bibfield  {title} {\enquote {\bibinfo {title}
  {{On relations between multi - gluon and multigraviton scattering}},}\ }\href
  {\doibase 10.1016/0370-2693(88)90813-1} {\bibfield  {journal} {\bibinfo
  {journal} {Phys. Lett. B}\ }\textbf {\bibinfo {volume} {211}},\ \bibinfo
  {pages} {91--94} (\bibinfo {year} {1988})}\BibitemShut {NoStop}%
\bibitem [{\citenamefont {Bern}\ \emph
  {et~al.}(2008{\natexlab{a}})\citenamefont {Bern}, \citenamefont {Carrasco},\
  and\ \citenamefont {Johansson}}]{Bern:2008qj}%
  \BibitemOpen
  \bibfield  {author} {\bibinfo {author} {\bibfnamefont {Z.}~\bibnamefont
  {Bern}}, \bibinfo {author} {\bibfnamefont {J.~J.~M.}\ \bibnamefont
  {Carrasco}}, \ and\ \bibinfo {author} {\bibfnamefont {Henrik}\ \bibnamefont
  {Johansson}},\ }\bibfield  {title} {\enquote {\bibinfo {title} {{New
  Relations for Gauge-Theory Amplitudes}},}\ }\href {\doibase
  10.1103/PhysRevD.78.085011} {\bibfield  {journal} {\bibinfo  {journal} {Phys.
  Rev.}\ }\textbf {\bibinfo {volume} {D78}},\ \bibinfo {pages} {085011}
  (\bibinfo {year} {2008}{\natexlab{a}})},\ \Eprint
  {http://arxiv.org/abs/0805.3993} {arXiv:0805.3993 [hep-ph]} \BibitemShut
  {NoStop}%
\bibitem [{\citenamefont {Bern}\ \emph {et~al.}(2010)\citenamefont {Bern},
  \citenamefont {Carrasco},\ and\ \citenamefont {Johansson}}]{Bern:2010ue}%
  \BibitemOpen
  \bibfield  {author} {\bibinfo {author} {\bibfnamefont {Zvi}\ \bibnamefont
  {Bern}}, \bibinfo {author} {\bibfnamefont {John Joseph~M.}\ \bibnamefont
  {Carrasco}}, \ and\ \bibinfo {author} {\bibfnamefont {Henrik}\ \bibnamefont
  {Johansson}},\ }\bibfield  {title} {\enquote {\bibinfo {title} {{Perturbative
  Quantum Gravity as a Double Copy of Gauge Theory}},}\ }\href {\doibase
  10.1103/PhysRevLett.105.061602} {\bibfield  {journal} {\bibinfo  {journal}
  {Phys. Rev. Lett.}\ }\textbf {\bibinfo {volume} {105}},\ \bibinfo {pages}
  {061602} (\bibinfo {year} {2010})},\ \Eprint {http://arxiv.org/abs/1004.0476}
  {arXiv:1004.0476 [hep-th]} \BibitemShut {NoStop}%
\bibitem [{\citenamefont {Bern}\ \emph
  {et~al.}(2019{\natexlab{a}})\citenamefont {Bern}, \citenamefont {Carrasco},
  \citenamefont {Chiodaroli}, \citenamefont {Johansson},\ and\ \citenamefont
  {Roiban}}]{Bern:2019prr}%
  \BibitemOpen
  \bibfield  {author} {\bibinfo {author} {\bibfnamefont {Zvi}\ \bibnamefont
  {Bern}}, \bibinfo {author} {\bibfnamefont {John~Joseph}\ \bibnamefont
  {Carrasco}}, \bibinfo {author} {\bibfnamefont {Marco}\ \bibnamefont
  {Chiodaroli}}, \bibinfo {author} {\bibfnamefont {Henrik}\ \bibnamefont
  {Johansson}}, \ and\ \bibinfo {author} {\bibfnamefont {Radu}\ \bibnamefont
  {Roiban}},\ }\bibfield  {title} {\enquote {\bibinfo {title} {{The Duality
  Between Color and Kinematics and its Applications}},}\ }\href@noop {} {\
  (\bibinfo {year} {2019}{\natexlab{a}})},\ \Eprint
  {http://arxiv.org/abs/1909.01358} {arXiv:1909.01358 [hep-th]} \BibitemShut
  {NoStop}%
\bibitem [{\citenamefont {Luna}\ \emph
  {et~al.}(2017{\natexlab{a}})\citenamefont {Luna}, \citenamefont {Monteiro},
  \citenamefont {Nicholson}, \citenamefont {Ochirov}, \citenamefont
  {O'Connell}, \citenamefont {Westerberg},\ and\ \citenamefont
  {White}}]{Luna:2016hge}%
  \BibitemOpen
  \bibfield  {author} {\bibinfo {author} {\bibfnamefont {Andrs}\ \bibnamefont
  {Luna}}, \bibinfo {author} {\bibfnamefont {Ricardo}\ \bibnamefont
  {Monteiro}}, \bibinfo {author} {\bibfnamefont {Isobel}\ \bibnamefont
  {Nicholson}}, \bibinfo {author} {\bibfnamefont {Alexander}\ \bibnamefont
  {Ochirov}}, \bibinfo {author} {\bibfnamefont {Donal}\ \bibnamefont
  {O'Connell}}, \bibinfo {author} {\bibfnamefont {Niclas}\ \bibnamefont
  {Westerberg}}, \ and\ \bibinfo {author} {\bibfnamefont {Chris~D.}\
  \bibnamefont {White}},\ }\bibfield  {title} {\enquote {\bibinfo {title}
  {{Perturbative spacetimes from Yang-Mills theory}},}\ }\href {\doibase
  10.1007/JHEP04(2017)069} {\bibfield  {journal} {\bibinfo  {journal} {JHEP}\
  }\textbf {\bibinfo {volume} {04}},\ \bibinfo {pages} {069} (\bibinfo {year}
  {2017}{\natexlab{a}})},\ \Eprint {http://arxiv.org/abs/1611.07508}
  {arXiv:1611.07508 [hep-th]} \BibitemShut {NoStop}%
\bibitem [{\citenamefont {Johansson}\ and\ \citenamefont
  {Ochirov}(2015)}]{Johansson:2014zca}%
  \BibitemOpen
  \bibfield  {author} {\bibinfo {author} {\bibfnamefont {Henrik}\ \bibnamefont
  {Johansson}}\ and\ \bibinfo {author} {\bibfnamefont {Alexander}\ \bibnamefont
  {Ochirov}},\ }\bibfield  {title} {\enquote {\bibinfo {title} {{Pure Gravities
  via Color-Kinematics Duality for Fundamental Matter}},}\ }\href {\doibase
  10.1007/JHEP11(2015)046} {\bibfield  {journal} {\bibinfo  {journal} {JHEP}\
  }\textbf {\bibinfo {volume} {11}},\ \bibinfo {pages} {046} (\bibinfo {year}
  {2015})},\ \Eprint {http://arxiv.org/abs/1407.4772} {arXiv:1407.4772
  [hep-th]} \BibitemShut {NoStop}%
\bibitem [{\citenamefont {Luna}\ \emph
  {et~al.}(2018{\natexlab{a}})\citenamefont {Luna}, \citenamefont {Nicholson},
  \citenamefont {O'Connell},\ and\ \citenamefont {White}}]{Luna:2017dtq}%
  \BibitemOpen
  \bibfield  {author} {\bibinfo {author} {\bibfnamefont {Andrs}\ \bibnamefont
  {Luna}}, \bibinfo {author} {\bibfnamefont {Isobel}\ \bibnamefont
  {Nicholson}}, \bibinfo {author} {\bibfnamefont {Donal}\ \bibnamefont
  {O'Connell}}, \ and\ \bibinfo {author} {\bibfnamefont {Chris~D.}\
  \bibnamefont {White}},\ }\bibfield  {title} {\enquote {\bibinfo {title}
  {{Inelastic Black Hole Scattering from Charged Scalar Amplitudes}},}\ }\href
  {\doibase 10.1007/JHEP03(2018)044} {\bibfield  {journal} {\bibinfo  {journal}
  {JHEP}\ }\textbf {\bibinfo {volume} {03}},\ \bibinfo {pages} {044} (\bibinfo
  {year} {2018}{\natexlab{a}})},\ \Eprint {http://arxiv.org/abs/1711.03901}
  {arXiv:1711.03901 [hep-th]} \BibitemShut {NoStop}%
\bibitem [{\citenamefont {Bern}\ \emph
  {et~al.}(2019{\natexlab{b}})\citenamefont {Bern}, \citenamefont {Cheung},
  \citenamefont {Roiban}, \citenamefont {Shen}, \citenamefont {Solon},\ and\
  \citenamefont {Zeng}}]{Bern:2019crd}%
  \BibitemOpen
  \bibfield  {author} {\bibinfo {author} {\bibfnamefont {Zvi}\ \bibnamefont
  {Bern}}, \bibinfo {author} {\bibfnamefont {Clifford}\ \bibnamefont {Cheung}},
  \bibinfo {author} {\bibfnamefont {Radu}\ \bibnamefont {Roiban}}, \bibinfo
  {author} {\bibfnamefont {Chia-Hsien}\ \bibnamefont {Shen}}, \bibinfo {author}
  {\bibfnamefont {Mikhail~P.}\ \bibnamefont {Solon}}, \ and\ \bibinfo {author}
  {\bibfnamefont {Mao}\ \bibnamefont {Zeng}},\ }\bibfield  {title} {\enquote
  {\bibinfo {title} {{Black Hole Binary Dynamics from the Double Copy and
  Effective Theory}},}\ }\href {\doibase 10.1007/JHEP10(2019)206} {\bibfield
  {journal} {\bibinfo  {journal} {JHEP}\ }\textbf {\bibinfo {volume} {10}},\
  \bibinfo {pages} {206} (\bibinfo {year} {2019}{\natexlab{b}})},\ \Eprint
  {http://arxiv.org/abs/1908.01493} {arXiv:1908.01493 [hep-th]} \BibitemShut
  {NoStop}%
\bibitem [{\citenamefont {Bern}\ \emph
  {et~al.}(2012{\natexlab{a}})\citenamefont {Bern}, \citenamefont {Carrasco},
  \citenamefont {Dixon}, \citenamefont {Johansson},\ and\ \citenamefont
  {Roiban}}]{SimplifyingBCJ}%
  \BibitemOpen
  \bibfield  {author} {\bibinfo {author} {\bibfnamefont {Z.}~\bibnamefont
  {Bern}}, \bibinfo {author} {\bibfnamefont {J.~J.~M.}\ \bibnamefont
  {Carrasco}}, \bibinfo {author} {\bibfnamefont {L.~J.}\ \bibnamefont {Dixon}},
  \bibinfo {author} {\bibfnamefont {H.}~\bibnamefont {Johansson}}, \ and\
  \bibinfo {author} {\bibfnamefont {R.}~\bibnamefont {Roiban}},\ }\bibfield
  {title} {\enquote {\bibinfo {title} {{Simplifying multiloop integrands and
  ultraviolet divergences of gauge theory and gravity amplitudes}},}\ }\href
  {\doibase 10.1103/PhysRevD.85.105014} {\bibfield  {journal} {\bibinfo
  {journal} {Phys. Rev.}\ }\textbf {\bibinfo {volume} {D85}},\ \bibinfo {pages}
  {105014} (\bibinfo {year} {2012}{\natexlab{a}})},\ \Eprint
  {http://arxiv.org/abs/1201.5366} {arXiv:1201.5366 [hep-th]} \BibitemShut
  {NoStop}%
\bibitem [{\citenamefont {Bern}\ \emph {et~al.}(2011)\citenamefont {Bern},
  \citenamefont {Boucher-Veronneau},\ and\ \citenamefont
  {Johansson}}]{N46Sugra}%
  \BibitemOpen
  \bibfield  {author} {\bibinfo {author} {\bibfnamefont {Z.}~\bibnamefont
  {Bern}}, \bibinfo {author} {\bibfnamefont {C.}~\bibnamefont
  {Boucher-Veronneau}}, \ and\ \bibinfo {author} {\bibfnamefont
  {H.}~\bibnamefont {Johansson}},\ }\bibfield  {title} {\enquote {\bibinfo
  {title} {{${\cal N} = 4$ supergravity amplitudes from gauge theory at one
  loop}},}\ }\href {\doibase 10.1103/PhysRevD.84.105035} {\bibfield  {journal}
  {\bibinfo  {journal} {Phys. Rev.}\ }\textbf {\bibinfo {volume} {D84}},\
  \bibinfo {pages} {105035} (\bibinfo {year} {2011})},\ \Eprint
  {http://arxiv.org/abs/1107.1935} {arXiv:1107.1935 [hep-th]} \BibitemShut
  {NoStop}%
\bibitem [{\citenamefont {Boucher-Veronneau}\ and\ \citenamefont
  {Dixon}(2011)}]{N46Sugra2}%
  \BibitemOpen
  \bibfield  {author} {\bibinfo {author} {\bibfnamefont {C.}~\bibnamefont
  {Boucher-Veronneau}}\ and\ \bibinfo {author} {\bibfnamefont {L.~J.}\
  \bibnamefont {Dixon}},\ }\bibfield  {title} {\enquote {\bibinfo {title}
  {{${\cal N}\ge 4$ supergravity amplitudes from gauge theory at two Loops}},}\
  }\href {\doibase 10.1007/JHEP12(2011)046} {\bibfield  {journal} {\bibinfo
  {journal} {JHEP}\ }\textbf {\bibinfo {volume} {12}},\ \bibinfo {pages} {046}
  (\bibinfo {year} {2011})},\ \Eprint {http://arxiv.org/abs/1110.1132}
  {arXiv:1110.1132 [hep-th]} \BibitemShut {NoStop}%
\bibitem [{\citenamefont {Bern}\ \emph
  {et~al.}(2012{\natexlab{b}})\citenamefont {Bern}, \citenamefont {Davies},
  \citenamefont {Dennen},\ and\ \citenamefont {Huang}}]{Bern:2012cd}%
  \BibitemOpen
  \bibfield  {author} {\bibinfo {author} {\bibfnamefont {Zvi}\ \bibnamefont
  {Bern}}, \bibinfo {author} {\bibfnamefont {Scott}\ \bibnamefont {Davies}},
  \bibinfo {author} {\bibfnamefont {Tristan}\ \bibnamefont {Dennen}}, \ and\
  \bibinfo {author} {\bibfnamefont {Yu-tin}\ \bibnamefont {Huang}},\ }\bibfield
   {title} {\enquote {\bibinfo {title} {{Absence of Three-Loop Four-Point
  Divergences in N=4 Supergravity}},}\ }\href {\doibase
  10.1103/PhysRevLett.108.201301} {\bibfield  {journal} {\bibinfo  {journal}
  {Phys. Rev. Lett.}\ }\textbf {\bibinfo {volume} {108}},\ \bibinfo {pages}
  {201301} (\bibinfo {year} {2012}{\natexlab{b}})},\ \Eprint
  {http://arxiv.org/abs/1202.3423} {arXiv:1202.3423 [hep-th]} \BibitemShut
  {NoStop}%
\bibitem [{\citenamefont {Bern}\ \emph
  {et~al.}(2012{\natexlab{c}})\citenamefont {Bern}, \citenamefont {Davies},
  \citenamefont {Dennen},\ and\ \citenamefont {Huang}}]{Bern:2012gh}%
  \BibitemOpen
  \bibfield  {author} {\bibinfo {author} {\bibfnamefont {Zvi}\ \bibnamefont
  {Bern}}, \bibinfo {author} {\bibfnamefont {Scott}\ \bibnamefont {Davies}},
  \bibinfo {author} {\bibfnamefont {Tristan}\ \bibnamefont {Dennen}}, \ and\
  \bibinfo {author} {\bibfnamefont {Yu-tin}\ \bibnamefont {Huang}},\ }\bibfield
   {title} {\enquote {\bibinfo {title} {{Ultraviolet cancellations in
  half-maximal supergravity as a consequence of the double-copy structure}},}\
  }\href {\doibase 10.1103/PhysRevD.86.105014} {\bibfield  {journal} {\bibinfo
  {journal} {Phys. Rev.}\ }\textbf {\bibinfo {volume} {D86}},\ \bibinfo {pages}
  {105014} (\bibinfo {year} {2012}{\natexlab{c}})},\ \Eprint
  {http://arxiv.org/abs/1209.2472} {arXiv:1209.2472 [hep-th]} \BibitemShut
  {NoStop}%
\bibitem [{\citenamefont {Boels}\ and\ \citenamefont
  {Isermann}(2013)}]{Boels:2012sy}%
  \BibitemOpen
  \bibfield  {author} {\bibinfo {author} {\bibfnamefont {Rutger~H.}\
  \bibnamefont {Boels}}\ and\ \bibinfo {author} {\bibfnamefont {Reinke~Sven}\
  \bibnamefont {Isermann}},\ }\bibfield  {title} {\enquote {\bibinfo {title}
  {{On powercounting in perturbative quantum gravity theories through
  color-kinematic duality}},}\ }\href {\doibase 10.1007/JHEP06(2013)017}
  {\bibfield  {journal} {\bibinfo  {journal} {JHEP}\ }\textbf {\bibinfo
  {volume} {06}},\ \bibinfo {pages} {017} (\bibinfo {year} {2013})},\ \Eprint
  {http://arxiv.org/abs/1212.3473} {arXiv:1212.3473 [hep-th]} \BibitemShut
  {NoStop}%
\bibitem [{\citenamefont {Bern}\ \emph {et~al.}(2013)\citenamefont {Bern},
  \citenamefont {Davies},\ and\ \citenamefont {Dennen}}]{Bern:2013qca}%
  \BibitemOpen
  \bibfield  {author} {\bibinfo {author} {\bibfnamefont {Zvi}\ \bibnamefont
  {Bern}}, \bibinfo {author} {\bibfnamefont {Scott}\ \bibnamefont {Davies}}, \
  and\ \bibinfo {author} {\bibfnamefont {Tristan}\ \bibnamefont {Dennen}},\
  }\bibfield  {title} {\enquote {\bibinfo {title} {{The ultraviolet structure
  of half-maximal supergravity with matter multiplets at two and three
  loops}},}\ }\href {\doibase 10.1103/PhysRevD.88.065007} {\bibfield  {journal}
  {\bibinfo  {journal} {Phys. Rev.}\ }\textbf {\bibinfo {volume} {D88}},\
  \bibinfo {pages} {065007} (\bibinfo {year} {2013})},\ \Eprint
  {http://arxiv.org/abs/1305.4876} {arXiv:1305.4876 [hep-th]} \BibitemShut
  {NoStop}%
\bibitem [{\citenamefont {Bern}\ \emph {et~al.}(2014)\citenamefont {Bern},
  \citenamefont {Davies},\ and\ \citenamefont {Dennen}}]{Bern:2014lha}%
  \BibitemOpen
  \bibfield  {author} {\bibinfo {author} {\bibfnamefont {Zvi}\ \bibnamefont
  {Bern}}, \bibinfo {author} {\bibfnamefont {Scott}\ \bibnamefont {Davies}}, \
  and\ \bibinfo {author} {\bibfnamefont {Tristan}\ \bibnamefont {Dennen}},\
  }\bibfield  {title} {\enquote {\bibinfo {title} {{The Ultraviolet Critical
  Dimension of Half-Maximal Supergravity at Three Loops}},}\ }\href@noop {} {\
  (\bibinfo {year} {2014})},\ \Eprint {http://arxiv.org/abs/1412.2441}
  {arXiv:1412.2441 [hep-th]} \BibitemShut {NoStop}%
\bibitem [{\citenamefont {Bern}\ \emph {et~al.}(2018)\citenamefont {Bern},
  \citenamefont {Carrasco}, \citenamefont {Chen}, \citenamefont {Edison},
  \citenamefont {Johansson}, \citenamefont {Parra-Martinez}, \citenamefont
  {Roiban},\ and\ \citenamefont {Zeng}}]{UVFiveLoops}%
  \BibitemOpen
  \bibfield  {author} {\bibinfo {author} {\bibfnamefont {Zvi}\ \bibnamefont
  {Bern}}, \bibinfo {author} {\bibfnamefont {John~Joseph}\ \bibnamefont
  {Carrasco}}, \bibinfo {author} {\bibfnamefont {Wei-Ming}\ \bibnamefont
  {Chen}}, \bibinfo {author} {\bibfnamefont {Alex}\ \bibnamefont {Edison}},
  \bibinfo {author} {\bibfnamefont {Henrik}\ \bibnamefont {Johansson}},
  \bibinfo {author} {\bibfnamefont {Julio}\ \bibnamefont {Parra-Martinez}},
  \bibinfo {author} {\bibfnamefont {Radu}\ \bibnamefont {Roiban}}, \ and\
  \bibinfo {author} {\bibfnamefont {Mao}\ \bibnamefont {Zeng}},\ }\bibfield
  {title} {\enquote {\bibinfo {title} {{Ultraviolet properties of $\mathcal N =
  8$ supergravity at five loops}},}\ }\href {\doibase
  10.1103/PhysRevD.98.086021} {\bibfield  {journal} {\bibinfo  {journal} {Phys.
  Rev.}\ }\textbf {\bibinfo {volume} {D98}},\ \bibinfo {pages} {086021}
  (\bibinfo {year} {2018})},\ \Eprint {http://arxiv.org/abs/1804.09311}
  {arXiv:1804.09311 [hep-th]} \BibitemShut {NoStop}%
\bibitem [{\citenamefont {Herrmann}\ and\ \citenamefont
  {Trnka}(2016)}]{Herrmann:2016qea}%
  \BibitemOpen
  \bibfield  {author} {\bibinfo {author} {\bibfnamefont {Enrico}\ \bibnamefont
  {Herrmann}}\ and\ \bibinfo {author} {\bibfnamefont {Jaroslav}\ \bibnamefont
  {Trnka}},\ }\bibfield  {title} {\enquote {\bibinfo {title} {{Gravity On-shell
  diagrams}},}\ }\href {\doibase 10.1007/JHEP11(2016)136} {\bibfield  {journal}
  {\bibinfo  {journal} {JHEP}\ }\textbf {\bibinfo {volume} {11}},\ \bibinfo
  {pages} {136} (\bibinfo {year} {2016})},\ \Eprint
  {http://arxiv.org/abs/1604.03479} {arXiv:1604.03479 [hep-th]} \BibitemShut
  {NoStop}%
\bibitem [{\citenamefont {Herrmann}\ and\ \citenamefont
  {Trnka}(2019)}]{HerrmannTrnkaUVGrav}%
  \BibitemOpen
  \bibfield  {author} {\bibinfo {author} {\bibfnamefont {Enrico}\ \bibnamefont
  {Herrmann}}\ and\ \bibinfo {author} {\bibfnamefont {Jaroslav}\ \bibnamefont
  {Trnka}},\ }\bibfield  {title} {\enquote {\bibinfo {title} {{UV cancellations
  in gravity loop integrands}},}\ }\href {\doibase 10.1007/JHEP02(2019)084}
  {\bibfield  {journal} {\bibinfo  {journal} {JHEP}\ }\textbf {\bibinfo
  {volume} {02}},\ \bibinfo {pages} {084} (\bibinfo {year} {2019})},\ \Eprint
  {http://arxiv.org/abs/1808.10446} {arXiv:1808.10446 [hep-th]} \BibitemShut
  {NoStop}%
\bibitem [{\citenamefont {Bern}\ \emph
  {et~al.}(2008{\natexlab{b}})\citenamefont {Bern}, \citenamefont {Carrasco},
  \citenamefont {Forde}, \citenamefont {Ita},\ and\ \citenamefont
  {Johansson}}]{Bern:2007xj}%
  \BibitemOpen
  \bibfield  {author} {\bibinfo {author} {\bibfnamefont {Z.}~\bibnamefont
  {Bern}}, \bibinfo {author} {\bibfnamefont {J.~J.}\ \bibnamefont {Carrasco}},
  \bibinfo {author} {\bibfnamefont {D.}~\bibnamefont {Forde}}, \bibinfo
  {author} {\bibfnamefont {H.}~\bibnamefont {Ita}}, \ and\ \bibinfo {author}
  {\bibfnamefont {Henrik}\ \bibnamefont {Johansson}},\ }\bibfield  {title}
  {\enquote {\bibinfo {title} {{Unexpected Cancellations in Gravity
  Theories}},}\ }\href {\doibase 10.1103/PhysRevD.77.025010} {\bibfield
  {journal} {\bibinfo  {journal} {Phys. Rev.}\ }\textbf {\bibinfo {volume}
  {D77}},\ \bibinfo {pages} {025010} (\bibinfo {year} {2008}{\natexlab{b}})},\
  \Eprint {http://arxiv.org/abs/0707.1035} {arXiv:0707.1035 [hep-th]}
  \BibitemShut {NoStop}%
\bibitem [{\citenamefont {Saotome}\ and\ \citenamefont
  {Akhoury}(2013)}]{Saotome2012vy}%
  \BibitemOpen
  \bibfield  {author} {\bibinfo {author} {\bibfnamefont {Ryo}\ \bibnamefont
  {Saotome}}\ and\ \bibinfo {author} {\bibfnamefont {Ratindranath}\
  \bibnamefont {Akhoury}},\ }\bibfield  {title} {\enquote {\bibinfo {title}
  {{Relationship between gravity and gauge scattering in the high energy
  limit}},}\ }\href {\doibase 10.1007/JHEP01(2013)123} {\bibfield  {journal}
  {\bibinfo  {journal} {JHEP}\ }\textbf {\bibinfo {volume} {01}},\ \bibinfo
  {pages} {123} (\bibinfo {year} {2013})},\ \Eprint
  {http://arxiv.org/abs/1210.8111} {arXiv:1210.8111 [hep-th]} \BibitemShut
  {NoStop}%
\bibitem [{\citenamefont {Monteiro}\ \emph {et~al.}(2014)\citenamefont
  {Monteiro}, \citenamefont {O'Connell},\ and\ \citenamefont
  {White}}]{Monteiro2014cda}%
  \BibitemOpen
  \bibfield  {author} {\bibinfo {author} {\bibfnamefont {Ricardo}\ \bibnamefont
  {Monteiro}}, \bibinfo {author} {\bibfnamefont {Donal}\ \bibnamefont
  {O'Connell}}, \ and\ \bibinfo {author} {\bibfnamefont {Chris~D.}\
  \bibnamefont {White}},\ }\bibfield  {title} {\enquote {\bibinfo {title}
  {{Black holes and the double copy}},}\ }\href {\doibase
  10.1007/JHEP12(2014)056} {\bibfield  {journal} {\bibinfo  {journal} {JHEP}\
  }\textbf {\bibinfo {volume} {12}},\ \bibinfo {pages} {056} (\bibinfo {year}
  {2014})},\ \Eprint {http://arxiv.org/abs/1410.0239} {arXiv:1410.0239
  [hep-th]} \BibitemShut {NoStop}%
\bibitem [{\citenamefont {Luna}\ \emph {et~al.}(2015)\citenamefont {Luna},
  \citenamefont {Monteiro}, \citenamefont {O'Connell},\ and\ \citenamefont
  {White}}]{Luna2015paa}%
  \BibitemOpen
  \bibfield  {author} {\bibinfo {author} {\bibfnamefont {Andr{\'e}s}\
  \bibnamefont {Luna}}, \bibinfo {author} {\bibfnamefont {Ricardo}\
  \bibnamefont {Monteiro}}, \bibinfo {author} {\bibfnamefont {Donal}\
  \bibnamefont {O'Connell}}, \ and\ \bibinfo {author} {\bibfnamefont
  {Chris~D.}\ \bibnamefont {White}},\ }\bibfield  {title} {\enquote {\bibinfo
  {title} {{The classical double copy for Taub--NUT spacetime}},}\ }\href
  {\doibase 10.1016/j.physletb.2015.09.021} {\bibfield  {journal} {\bibinfo
  {journal} {Phys. Lett.}\ }\textbf {\bibinfo {volume} {B750}},\ \bibinfo
  {pages} {272--277} (\bibinfo {year} {2015})},\ \Eprint
  {http://arxiv.org/abs/1507.01869} {arXiv:1507.01869 [hep-th]} \BibitemShut
  {NoStop}%
\bibitem [{\citenamefont {Ridgway}\ and\ \citenamefont
  {Wise}(2016)}]{Ridgway2015fdl}%
  \BibitemOpen
  \bibfield  {author} {\bibinfo {author} {\bibfnamefont {Alexander~K.}\
  \bibnamefont {Ridgway}}\ and\ \bibinfo {author} {\bibfnamefont {Mark~B.}\
  \bibnamefont {Wise}},\ }\bibfield  {title} {\enquote {\bibinfo {title}
  {{Static spherically symmetric Kerr-Schild metrics and implications for the
  classical double copy}},}\ }\href {\doibase 10.1103/PhysRevD.94.044023}
  {\bibfield  {journal} {\bibinfo  {journal} {Phys. Rev.}\ }\textbf {\bibinfo
  {volume} {D94}},\ \bibinfo {pages} {044023} (\bibinfo {year} {2016})},\
  \Eprint {http://arxiv.org/abs/1512.02243} {arXiv:1512.02243 [hep-th]}
  \BibitemShut {NoStop}%
\bibitem [{\citenamefont {Luna}\ \emph {et~al.}(2016)\citenamefont {Luna},
  \citenamefont {Monteiro}, \citenamefont {Nicholson}, \citenamefont
  {O'Connell},\ and\ \citenamefont {White}}]{Luna2016due}%
  \BibitemOpen
  \bibfield  {author} {\bibinfo {author} {\bibfnamefont {Andr{\'e}s}\
  \bibnamefont {Luna}}, \bibinfo {author} {\bibfnamefont {Ricardo}\
  \bibnamefont {Monteiro}}, \bibinfo {author} {\bibfnamefont {Isobel}\
  \bibnamefont {Nicholson}}, \bibinfo {author} {\bibfnamefont {Donal}\
  \bibnamefont {O'Connell}}, \ and\ \bibinfo {author} {\bibfnamefont
  {Chris~D.}\ \bibnamefont {White}},\ }\bibfield  {title} {\enquote {\bibinfo
  {title} {{The double copy: Bremsstrahlung and accelerating black holes}},}\
  }\href {\doibase 10.1007/JHEP06(2016)023} {\bibfield  {journal} {\bibinfo
  {journal} {JHEP}\ }\textbf {\bibinfo {volume} {06}},\ \bibinfo {pages} {023}
  (\bibinfo {year} {2016})},\ \Eprint {http://arxiv.org/abs/1603.05737}
  {arXiv:1603.05737 [hep-th]} \BibitemShut {NoStop}%
\bibitem [{\citenamefont {White}(2016)}]{White2016jzc}%
  \BibitemOpen
  \bibfield  {author} {\bibinfo {author} {\bibfnamefont {Chris~D.}\
  \bibnamefont {White}},\ }\bibfield  {title} {\enquote {\bibinfo {title}
  {{Exact solutions for the biadjoint scalar field}},}\ }\href {\doibase
  10.1016/j.physletb.2016.10.052} {\bibfield  {journal} {\bibinfo  {journal}
  {Phys. Lett.}\ }\textbf {\bibinfo {volume} {B763}},\ \bibinfo {pages}
  {365--369} (\bibinfo {year} {2016})},\ \Eprint
  {http://arxiv.org/abs/1606.04724} {arXiv:1606.04724 [hep-th]} \BibitemShut
  {NoStop}%
\bibitem [{\citenamefont {Goldberger}\ and\ \citenamefont
  {Ridgway}(2017)}]{Goldberger2016iau}%
  \BibitemOpen
  \bibfield  {author} {\bibinfo {author} {\bibfnamefont {Walter~D.}\
  \bibnamefont {Goldberger}}\ and\ \bibinfo {author} {\bibfnamefont
  {Alexander~K.}\ \bibnamefont {Ridgway}},\ }\bibfield  {title} {\enquote
  {\bibinfo {title} {{Radiation and the classical double copy for color
  charges}},}\ }\href {\doibase 10.1103/PhysRevD.95.125010} {\bibfield
  {journal} {\bibinfo  {journal} {Phys. Rev.}\ }\textbf {\bibinfo {volume}
  {D95}},\ \bibinfo {pages} {125010} (\bibinfo {year} {2017})},\ \Eprint
  {http://arxiv.org/abs/1611.03493} {arXiv:1611.03493 [hep-th]} \BibitemShut
  {NoStop}%
\bibitem [{\citenamefont {Cardoso}\ \emph {et~al.}(2017)\citenamefont
  {Cardoso}, \citenamefont {Nagy},\ and\ \citenamefont
  {Nampuri}}]{Cardoso2016amd}%
  \BibitemOpen
  \bibfield  {author} {\bibinfo {author} {\bibfnamefont {Gabriel}\ \bibnamefont
  {Cardoso}}, \bibinfo {author} {\bibfnamefont {Silvia}\ \bibnamefont {Nagy}},
  \ and\ \bibinfo {author} {\bibfnamefont {Suresh}\ \bibnamefont {Nampuri}},\
  }\bibfield  {title} {\enquote {\bibinfo {title} {{Multi-centered $
  \mathcal{N}=2 $ BPS black holes: a double copy description}},}\ }\href
  {\doibase 10.1007/JHEP04(2017)037} {\bibfield  {journal} {\bibinfo  {journal}
  {JHEP}\ }\textbf {\bibinfo {volume} {04}},\ \bibinfo {pages} {037} (\bibinfo
  {year} {2017})},\ \Eprint {http://arxiv.org/abs/1611.04409} {arXiv:1611.04409
  [hep-th]} \BibitemShut {NoStop}%
\bibitem [{\citenamefont {Luna}\ \emph
  {et~al.}(2017{\natexlab{b}})\citenamefont {Luna}, \citenamefont {Monteiro},
  \citenamefont {Nicholson}, \citenamefont {Ochirov}, \citenamefont
  {O'Connell}, \citenamefont {Westerberg},\ and\ \citenamefont
  {White}}]{Luna2016hge}%
  \BibitemOpen
  \bibfield  {author} {\bibinfo {author} {\bibfnamefont {Andr{\'e}s}\
  \bibnamefont {Luna}}, \bibinfo {author} {\bibfnamefont {Ricardo}\
  \bibnamefont {Monteiro}}, \bibinfo {author} {\bibfnamefont {Isobel}\
  \bibnamefont {Nicholson}}, \bibinfo {author} {\bibfnamefont {Alexander}\
  \bibnamefont {Ochirov}}, \bibinfo {author} {\bibfnamefont {Donal}\
  \bibnamefont {O'Connell}}, \bibinfo {author} {\bibfnamefont {Niclas}\
  \bibnamefont {Westerberg}}, \ and\ \bibinfo {author} {\bibfnamefont
  {Chris~D.}\ \bibnamefont {White}},\ }\bibfield  {title} {\enquote {\bibinfo
  {title} {{Perturbative spacetimes from Yang-Mills theory}},}\ }\href
  {\doibase 10.1007/JHEP04(2017)069} {\bibfield  {journal} {\bibinfo  {journal}
  {JHEP}\ }\textbf {\bibinfo {volume} {04}},\ \bibinfo {pages} {069} (\bibinfo
  {year} {2017}{\natexlab{b}})},\ \Eprint {http://arxiv.org/abs/1611.07508}
  {arXiv:1611.07508 [hep-th]} \BibitemShut {NoStop}%
\bibitem [{\citenamefont {Goldberger}\ \emph {et~al.}(2017)\citenamefont
  {Goldberger}, \citenamefont {Prabhu},\ and\ \citenamefont
  {Thompson}}]{Goldberger2017frp}%
  \BibitemOpen
  \bibfield  {author} {\bibinfo {author} {\bibfnamefont {Walter~D.}\
  \bibnamefont {Goldberger}}, \bibinfo {author} {\bibfnamefont {Siddharth~G.}\
  \bibnamefont {Prabhu}}, \ and\ \bibinfo {author} {\bibfnamefont
  {Jedidiah~O.}\ \bibnamefont {Thompson}},\ }\bibfield  {title} {\enquote
  {\bibinfo {title} {{Classical gluon and graviton radiation from the
  bi-adjoint scalar double copy}},}\ }\href {\doibase
  10.1103/PhysRevD.96.065009} {\bibfield  {journal} {\bibinfo  {journal} {Phys.
  Rev.}\ }\textbf {\bibinfo {volume} {D96}},\ \bibinfo {pages} {065009}
  (\bibinfo {year} {2017})},\ \Eprint {http://arxiv.org/abs/1705.09263}
  {arXiv:1705.09263 [hep-th]} \BibitemShut {NoStop}%
\bibitem [{\citenamefont {Adamo}\ \emph {et~al.}(2018)\citenamefont {Adamo},
  \citenamefont {Casali}, \citenamefont {Mason},\ and\ \citenamefont
  {Nekovar}}]{Adamo2017nia}%
  \BibitemOpen
  \bibfield  {author} {\bibinfo {author} {\bibfnamefont {Tim}\ \bibnamefont
  {Adamo}}, \bibinfo {author} {\bibfnamefont {Eduardo}\ \bibnamefont {Casali}},
  \bibinfo {author} {\bibfnamefont {Lionel}\ \bibnamefont {Mason}}, \ and\
  \bibinfo {author} {\bibfnamefont {Stefan}\ \bibnamefont {Nekovar}},\
  }\bibfield  {title} {\enquote {\bibinfo {title} {{Scattering on plane waves
  and the double copy}},}\ }\href {\doibase 10.1088/1361-6382/aa9961}
  {\bibfield  {journal} {\bibinfo  {journal} {Class. Quant. Grav.}\ }\textbf
  {\bibinfo {volume} {35}},\ \bibinfo {pages} {015004} (\bibinfo {year}
  {2018})},\ \Eprint {http://arxiv.org/abs/1706.08925} {arXiv:1706.08925
  [hep-th]} \BibitemShut {NoStop}%
\bibitem [{\citenamefont {De~Smet}\ and\ \citenamefont
  {White}(2017)}]{DeSmet2017rve}%
  \BibitemOpen
  \bibfield  {author} {\bibinfo {author} {\bibfnamefont {Pieter-Jan}\
  \bibnamefont {De~Smet}}\ and\ \bibinfo {author} {\bibfnamefont {Chris~D.}\
  \bibnamefont {White}},\ }\bibfield  {title} {\enquote {\bibinfo {title}
  {{Extended solutions for the biadjoint scalar field}},}\ }\href {\doibase
  10.1016/j.physletb.2017.11.007} {\bibfield  {journal} {\bibinfo  {journal}
  {Phys. Lett.}\ }\textbf {\bibinfo {volume} {B775}},\ \bibinfo {pages}
  {163--167} (\bibinfo {year} {2017})},\ \Eprint
  {http://arxiv.org/abs/1708.01103} {arXiv:1708.01103 [hep-th]} \BibitemShut
  {NoStop}%
\bibitem [{\citenamefont {Bahjat-Abbas}\ \emph {et~al.}(2017)\citenamefont
  {Bahjat-Abbas}, \citenamefont {Luna},\ and\ \citenamefont
  {White}}]{BahjatAbbas2017htu}%
  \BibitemOpen
  \bibfield  {author} {\bibinfo {author} {\bibfnamefont {Nadia}\ \bibnamefont
  {Bahjat-Abbas}}, \bibinfo {author} {\bibfnamefont {Andr{\'e}s}\ \bibnamefont
  {Luna}}, \ and\ \bibinfo {author} {\bibfnamefont {Chris~D.}\ \bibnamefont
  {White}},\ }\bibfield  {title} {\enquote {\bibinfo {title} {{The Kerr-Schild
  double copy in curved spacetime}},}\ }\href {\doibase
  10.1007/JHEP12(2017)004} {\bibfield  {journal} {\bibinfo  {journal} {JHEP}\
  }\textbf {\bibinfo {volume} {12}},\ \bibinfo {pages} {004} (\bibinfo {year}
  {2017})},\ \Eprint {http://arxiv.org/abs/1710.01953} {arXiv:1710.01953
  [hep-th]} \BibitemShut {NoStop}%
\bibitem [{\citenamefont {Carrillo-Gonz{\'a}lez}\ \emph
  {et~al.}(2018)\citenamefont {Carrillo-Gonz{\'a}lez}, \citenamefont {Penco},\
  and\ \citenamefont {Trodden}}]{CarrilloGonzalez2017iyj}%
  \BibitemOpen
  \bibfield  {author} {\bibinfo {author} {\bibfnamefont {Mariana}\ \bibnamefont
  {Carrillo-Gonz{\'a}lez}}, \bibinfo {author} {\bibfnamefont {Riccardo}\
  \bibnamefont {Penco}}, \ and\ \bibinfo {author} {\bibfnamefont {Mark}\
  \bibnamefont {Trodden}},\ }\bibfield  {title} {\enquote {\bibinfo {title}
  {{The classical double copy in maximally symmetric spacetimes}},}\ }\href
  {\doibase 10.1007/JHEP04(2018)028} {\bibfield  {journal} {\bibinfo  {journal}
  {JHEP}\ }\textbf {\bibinfo {volume} {04}},\ \bibinfo {pages} {028} (\bibinfo
  {year} {2018})},\ \Eprint {http://arxiv.org/abs/1711.01296} {arXiv:1711.01296
  [hep-th]} \BibitemShut {NoStop}%
\bibitem [{\citenamefont {Goldberger}\ \emph {et~al.}(2018)\citenamefont
  {Goldberger}, \citenamefont {Li},\ and\ \citenamefont
  {Prabhu}}]{Goldberger2017ogt}%
  \BibitemOpen
  \bibfield  {author} {\bibinfo {author} {\bibfnamefont {Walter~D.}\
  \bibnamefont {Goldberger}}, \bibinfo {author} {\bibfnamefont {Jingping}\
  \bibnamefont {Li}}, \ and\ \bibinfo {author} {\bibfnamefont {Siddharth~G.}\
  \bibnamefont {Prabhu}},\ }\bibfield  {title} {\enquote {\bibinfo {title}
  {{Spinning particles, axion radiation, and the classical double copy}},}\
  }\href {\doibase 10.1103/PhysRevD.97.105018} {\bibfield  {journal} {\bibinfo
  {journal} {Phys. Rev.}\ }\textbf {\bibinfo {volume} {D97}},\ \bibinfo {pages}
  {105018} (\bibinfo {year} {2018})},\ \Eprint
  {http://arxiv.org/abs/1712.09250} {arXiv:1712.09250 [hep-th]} \BibitemShut
  {NoStop}%
\bibitem [{\citenamefont {Li}\ and\ \citenamefont {Prabhu}(2018)}]{Li2018qap}%
  \BibitemOpen
  \bibfield  {author} {\bibinfo {author} {\bibfnamefont {Jingping}\
  \bibnamefont {Li}}\ and\ \bibinfo {author} {\bibfnamefont {Siddharth~G.}\
  \bibnamefont {Prabhu}},\ }\bibfield  {title} {\enquote {\bibinfo {title}
  {{Gravitational radiation from the classical spinning double copy}},}\ }\href
  {\doibase 10.1103/PhysRevD.97.105019} {\bibfield  {journal} {\bibinfo
  {journal} {Phys. Rev.}\ }\textbf {\bibinfo {volume} {D97}},\ \bibinfo {pages}
  {105019} (\bibinfo {year} {2018})},\ \Eprint
  {http://arxiv.org/abs/1803.02405} {arXiv:1803.02405 [hep-th]} \BibitemShut
  {NoStop}%
\bibitem [{\citenamefont {Ilderton}(2018)}]{Ilderton:2018lsf}%
  \BibitemOpen
  \bibfield  {author} {\bibinfo {author} {\bibfnamefont {A.}~\bibnamefont
  {Ilderton}},\ }\bibfield  {title} {\enquote {\bibinfo {title}
  {{Screw-symmetric gravitational waves: a double copy of the vortex}},}\
  }\href {\doibase 10.1016/j.physletb.2018.04.069} {\bibfield  {journal}
  {\bibinfo  {journal} {Phys. Lett.}\ }\textbf {\bibinfo {volume} {B782}},\
  \bibinfo {pages} {22--27} (\bibinfo {year} {2018})},\ \Eprint
  {http://arxiv.org/abs/1804.07290} {arXiv:1804.07290 [gr-qc]} \BibitemShut
  {NoStop}%
\bibitem [{\citenamefont {Shen}(2018)}]{ShenWorldLine}%
  \BibitemOpen
  \bibfield  {author} {\bibinfo {author} {\bibfnamefont {Chia-Hsien}\
  \bibnamefont {Shen}},\ }\bibfield  {title} {\enquote {\bibinfo {title}
  {{Gravitational radiation from color-kinematics duality}},}\ }\href {\doibase
  10.1007/JHEP11(2018)162} {\bibfield  {journal} {\bibinfo  {journal} {JHEP}\
  }\textbf {\bibinfo {volume} {11}},\ \bibinfo {pages} {162} (\bibinfo {year}
  {2018})},\ \Eprint {http://arxiv.org/abs/1806.07388} {arXiv:1806.07388
  [hep-th]} \BibitemShut {NoStop}%
\bibitem [{\citenamefont {Lee}(2018)}]{Lee:2018gxc}%
  \BibitemOpen
  \bibfield  {author} {\bibinfo {author} {\bibfnamefont {Kanghoon}\
  \bibnamefont {Lee}},\ }\bibfield  {title} {\enquote {\bibinfo {title}
  {{Kerr-Schild double field theory and classical double copy}},}\ }\href
  {\doibase 10.1007/JHEP10(2018)027} {\bibfield  {journal} {\bibinfo  {journal}
  {JHEP}\ }\textbf {\bibinfo {volume} {10}},\ \bibinfo {pages} {027} (\bibinfo
  {year} {2018})},\ \Eprint {http://arxiv.org/abs/1807.08443} {arXiv:1807.08443
  [hep-th]} \BibitemShut {NoStop}%
\bibitem [{\citenamefont {Plefka}\ \emph {et~al.}(2019)\citenamefont {Plefka},
  \citenamefont {Steinhoff},\ and\ \citenamefont
  {Wormsbecher}}]{Plefka:2018dpa}%
  \BibitemOpen
  \bibfield  {author} {\bibinfo {author} {\bibfnamefont {Jan}\ \bibnamefont
  {Plefka}}, \bibinfo {author} {\bibfnamefont {Jan}\ \bibnamefont {Steinhoff}},
  \ and\ \bibinfo {author} {\bibfnamefont {Wadim}\ \bibnamefont
  {Wormsbecher}},\ }\bibfield  {title} {\enquote {\bibinfo {title} {{Effective
  action of dilaton gravity as the classical double copy of Yang-Mills
  theory}},}\ }\href {\doibase 10.1103/PhysRevD.99.024021} {\bibfield
  {journal} {\bibinfo  {journal} {Phys. Rev.}\ }\textbf {\bibinfo {volume}
  {D99}},\ \bibinfo {pages} {024021} (\bibinfo {year} {2019})},\ \Eprint
  {http://arxiv.org/abs/1807.09859} {arXiv:1807.09859 [hep-th]} \BibitemShut
  {NoStop}%
\bibitem [{\citenamefont {Cheung}\ \emph {et~al.}(2018)\citenamefont {Cheung},
  \citenamefont {Rothstein},\ and\ \citenamefont {Solon}}]{CheungPM}%
  \BibitemOpen
  \bibfield  {author} {\bibinfo {author} {\bibfnamefont {Clifford}\
  \bibnamefont {Cheung}}, \bibinfo {author} {\bibfnamefont {Ira~Z.}\
  \bibnamefont {Rothstein}}, \ and\ \bibinfo {author} {\bibfnamefont
  {Mikhail~P.}\ \bibnamefont {Solon}},\ }\bibfield  {title} {\enquote {\bibinfo
  {title} {{From scattering amplitudes to classical potentials in the
  post-Minkowskian expansion}},}\ }\href {\doibase
  10.1103/PhysRevLett.121.251101} {\bibfield  {journal} {\bibinfo  {journal}
  {Phys. Rev. Lett.}\ }\textbf {\bibinfo {volume} {121}},\ \bibinfo {pages}
  {251101} (\bibinfo {year} {2018})},\ \Eprint
  {http://arxiv.org/abs/1808.02489} {arXiv:1808.02489 [hep-th]} \BibitemShut
  {NoStop}%
\bibitem [{\citenamefont {Berman}\ \emph {et~al.}(2019)\citenamefont {Berman},
  \citenamefont {Chac{\'o}n}, \citenamefont {Luna},\ and\ \citenamefont
  {White}}]{Berman:2018hwd}%
  \BibitemOpen
  \bibfield  {author} {\bibinfo {author} {\bibfnamefont {David~S.}\
  \bibnamefont {Berman}}, \bibinfo {author} {\bibfnamefont {Erick}\
  \bibnamefont {Chac{\'o}n}}, \bibinfo {author} {\bibfnamefont {Andr{\'e}s}\
  \bibnamefont {Luna}}, \ and\ \bibinfo {author} {\bibfnamefont {Chris~D.}\
  \bibnamefont {White}},\ }\bibfield  {title} {\enquote {\bibinfo {title} {{The
  self-dual classical double copy, and the Eguchi-Hanson instanton}},}\ }\href
  {\doibase 10.1007/JHEP01(2019)107} {\bibfield  {journal} {\bibinfo  {journal}
  {JHEP}\ }\textbf {\bibinfo {volume} {01}},\ \bibinfo {pages} {107} (\bibinfo
  {year} {2019})},\ \Eprint {http://arxiv.org/abs/1809.04063} {arXiv:1809.04063
  [hep-th]} \BibitemShut {NoStop}%
\bibitem [{\citenamefont {Gurses}\ and\ \citenamefont
  {Tekin}(2018)}]{Gurses:2018ckx}%
  \BibitemOpen
  \bibfield  {author} {\bibinfo {author} {\bibfnamefont {Metin}\ \bibnamefont
  {Gurses}}\ and\ \bibinfo {author} {\bibfnamefont {Bayram}\ \bibnamefont
  {Tekin}},\ }\bibfield  {title} {\enquote {\bibinfo {title} {{Classical double
  copy: Kerr-Schild-Kundt metrics from Yang-Mills theory}},}\ }\href {\doibase
  10.1103/PhysRevD.98.126017} {\bibfield  {journal} {\bibinfo  {journal} {Phys.
  Rev.}\ }\textbf {\bibinfo {volume} {D98}},\ \bibinfo {pages} {126017}
  (\bibinfo {year} {2018})},\ \Eprint {http://arxiv.org/abs/1810.03411}
  {arXiv:1810.03411 [gr-qc]} \BibitemShut {NoStop}%
\bibitem [{\citenamefont {Adamo}\ \emph {et~al.}(2019)\citenamefont {Adamo},
  \citenamefont {Casali}, \citenamefont {Mason},\ and\ \citenamefont
  {Nekovar}}]{Adamo:2018mpq}%
  \BibitemOpen
  \bibfield  {author} {\bibinfo {author} {\bibfnamefont {Tim}\ \bibnamefont
  {Adamo}}, \bibinfo {author} {\bibfnamefont {Eduardo}\ \bibnamefont {Casali}},
  \bibinfo {author} {\bibfnamefont {Lionel}\ \bibnamefont {Mason}}, \ and\
  \bibinfo {author} {\bibfnamefont {Stefan}\ \bibnamefont {Nekovar}},\
  }\bibfield  {title} {\enquote {\bibinfo {title} {{Plane wave backgrounds and
  colour-kinematics duality}},}\ }\href {\doibase 10.1007/JHEP02(2019)198}
  {\bibfield  {journal} {\bibinfo  {journal} {JHEP}\ }\textbf {\bibinfo
  {volume} {02}},\ \bibinfo {pages} {198} (\bibinfo {year} {2019})},\ \Eprint
  {http://arxiv.org/abs/1810.05115} {arXiv:1810.05115 [hep-th]} \BibitemShut
  {NoStop}%
\bibitem [{\citenamefont {Bahjat-Abbas}\ \emph {et~al.}(2019)\citenamefont
  {Bahjat-Abbas}, \citenamefont {Stark-Much{\~a}o},\ and\ \citenamefont
  {White}}]{Bahjat-Abbas:2018vgo}%
  \BibitemOpen
  \bibfield  {author} {\bibinfo {author} {\bibfnamefont {Nadia}\ \bibnamefont
  {Bahjat-Abbas}}, \bibinfo {author} {\bibfnamefont {Ricardo}\ \bibnamefont
  {Stark-Much{\~a}o}}, \ and\ \bibinfo {author} {\bibfnamefont {Chris~D.}\
  \bibnamefont {White}},\ }\bibfield  {title} {\enquote {\bibinfo {title}
  {{Biadjoint wires}},}\ }\href {\doibase 10.1016/j.physletb.2018.11.026}
  {\bibfield  {journal} {\bibinfo  {journal} {Phys. Lett.}\ }\textbf {\bibinfo
  {volume} {B788}},\ \bibinfo {pages} {274--279} (\bibinfo {year} {2019})},\
  \Eprint {http://arxiv.org/abs/1810.08118} {arXiv:1810.08118 [hep-th]}
  \BibitemShut {NoStop}%
\bibitem [{\citenamefont {Luna}\ \emph {et~al.}(2019)\citenamefont {Luna},
  \citenamefont {Monteiro}, \citenamefont {Nicholson},\ and\ \citenamefont
  {O'Connell}}]{Luna:2018dpt}%
  \BibitemOpen
  \bibfield  {author} {\bibinfo {author} {\bibfnamefont {Andr\'es}\
  \bibnamefont {Luna}}, \bibinfo {author} {\bibfnamefont {Ricardo}\
  \bibnamefont {Monteiro}}, \bibinfo {author} {\bibfnamefont {Isobel}\
  \bibnamefont {Nicholson}}, \ and\ \bibinfo {author} {\bibfnamefont {Donal}\
  \bibnamefont {O'Connell}},\ }\bibfield  {title} {\enquote {\bibinfo {title}
  {{Type D Spacetimes and the Weyl Double Copy}},}\ }\href {\doibase
  10.1088/1361-6382/ab03e6} {\bibfield  {journal} {\bibinfo  {journal} {Class.
  Quant. Grav.}\ }\textbf {\bibinfo {volume} {36}},\ \bibinfo {pages} {065003}
  (\bibinfo {year} {2019})},\ \Eprint {http://arxiv.org/abs/1810.08183}
  {arXiv:1810.08183 [hep-th]} \BibitemShut {NoStop}%
\bibitem [{\citenamefont {Kosower}\ \emph {et~al.}(2019)\citenamefont
  {Kosower}, \citenamefont {Maybee},\ and\ \citenamefont
  {O'Connell}}]{Kosower:2018adc}%
  \BibitemOpen
  \bibfield  {author} {\bibinfo {author} {\bibfnamefont {David~A.}\
  \bibnamefont {Kosower}}, \bibinfo {author} {\bibfnamefont {Ben}\ \bibnamefont
  {Maybee}}, \ and\ \bibinfo {author} {\bibfnamefont {Donal}\ \bibnamefont
  {O'Connell}},\ }\bibfield  {title} {\enquote {\bibinfo {title} {{Amplitudes,
  Observables, and Classical Scattering}},}\ }\href {\doibase
  10.1007/JHEP02(2019)137} {\bibfield  {journal} {\bibinfo  {journal} {JHEP}\
  }\textbf {\bibinfo {volume} {02}},\ \bibinfo {pages} {137} (\bibinfo {year}
  {2019})},\ \Eprint {http://arxiv.org/abs/1811.10950} {arXiv:1811.10950
  [hep-th]} \BibitemShut {NoStop}%
\bibitem [{\citenamefont {Farrow}\ \emph {et~al.}(2019)\citenamefont {Farrow},
  \citenamefont {Lipstein},\ and\ \citenamefont {McFadden}}]{Farrow:2018yni}%
  \BibitemOpen
  \bibfield  {author} {\bibinfo {author} {\bibfnamefont {Joseph~A.}\
  \bibnamefont {Farrow}}, \bibinfo {author} {\bibfnamefont {Arthur~E.}\
  \bibnamefont {Lipstein}}, \ and\ \bibinfo {author} {\bibfnamefont {Paul}\
  \bibnamefont {McFadden}},\ }\bibfield  {title} {\enquote {\bibinfo {title}
  {{Double copy structure of CFT correlators}},}\ }\href {\doibase
  10.1007/JHEP02(2019)130} {\bibfield  {journal} {\bibinfo  {journal} {JHEP}\
  }\textbf {\bibinfo {volume} {02}},\ \bibinfo {pages} {130} (\bibinfo {year}
  {2019})},\ \Eprint {http://arxiv.org/abs/1812.11129} {arXiv:1812.11129
  [hep-th]} \BibitemShut {NoStop}%
\bibitem [{\citenamefont {Bern}\ \emph
  {et~al.}(2019{\natexlab{c}})\citenamefont {Bern}, \citenamefont {Cheung},
  \citenamefont {Roiban}, \citenamefont {Shen}, \citenamefont {Solon},\ and\
  \citenamefont {Zeng}}]{Bern:2019nnu}%
  \BibitemOpen
  \bibfield  {author} {\bibinfo {author} {\bibfnamefont {Zvi}\ \bibnamefont
  {Bern}}, \bibinfo {author} {\bibfnamefont {Clifford}\ \bibnamefont {Cheung}},
  \bibinfo {author} {\bibfnamefont {Radu}\ \bibnamefont {Roiban}}, \bibinfo
  {author} {\bibfnamefont {Chia-Hsien}\ \bibnamefont {Shen}}, \bibinfo {author}
  {\bibfnamefont {Mikhail~P.}\ \bibnamefont {Solon}}, \ and\ \bibinfo {author}
  {\bibfnamefont {Mao}\ \bibnamefont {Zeng}},\ }\bibfield  {title} {\enquote
  {\bibinfo {title} {{Scattering Amplitudes and the Conservative Hamiltonian
  for Binary Systems at Third Post-Minkowskian Order}},}\ }\href {\doibase
  10.1103/PhysRevLett.122.201603} {\bibfield  {journal} {\bibinfo  {journal}
  {Phys. Rev. Lett.}\ }\textbf {\bibinfo {volume} {122}},\ \bibinfo {pages}
  {201603} (\bibinfo {year} {2019}{\natexlab{c}})},\ \Eprint
  {http://arxiv.org/abs/1901.04424} {arXiv:1901.04424 [hep-th]} \BibitemShut
  {NoStop}%
\bibitem [{\citenamefont {Antonelli}\ \emph {et~al.}(2019)\citenamefont
  {Antonelli}, \citenamefont {Buonanno}, \citenamefont {Steinhoff},
  \citenamefont {van~de Meent},\ and\ \citenamefont
  {Vines}}]{Antonelli:2019ytb}%
  \BibitemOpen
  \bibfield  {author} {\bibinfo {author} {\bibfnamefont {Andrea}\ \bibnamefont
  {Antonelli}}, \bibinfo {author} {\bibfnamefont {Alessandra}\ \bibnamefont
  {Buonanno}}, \bibinfo {author} {\bibfnamefont {Jan}\ \bibnamefont
  {Steinhoff}}, \bibinfo {author} {\bibfnamefont {Maarten}\ \bibnamefont
  {van~de Meent}}, \ and\ \bibinfo {author} {\bibfnamefont {Justin}\
  \bibnamefont {Vines}},\ }\bibfield  {title} {\enquote {\bibinfo {title}
  {{Energetics of two-body Hamiltonians in post-Minkowskian gravity}},}\ }\href
  {\doibase 10.1103/PhysRevD.99.104004} {\bibfield  {journal} {\bibinfo
  {journal} {Phys. Rev. D}\ }\textbf {\bibinfo {volume} {99}},\ \bibinfo
  {pages} {104004} (\bibinfo {year} {2019})},\ \Eprint
  {http://arxiv.org/abs/1901.07102} {arXiv:1901.07102 [gr-qc]} \BibitemShut
  {NoStop}%
\bibitem [{\citenamefont {Carrillo~Gonz{\'a}lez}\ \emph
  {et~al.}(2019)\citenamefont {Carrillo~Gonz{\'a}lez}, \citenamefont {Melcher},
  \citenamefont {Ratliff}, \citenamefont {Watson},\ and\ \citenamefont
  {White}}]{CarrilloGonzalez:2019gof}%
  \BibitemOpen
  \bibfield  {author} {\bibinfo {author} {\bibfnamefont {Mariana}\ \bibnamefont
  {Carrillo~Gonz{\'a}lez}}, \bibinfo {author} {\bibfnamefont {Brandon}\
  \bibnamefont {Melcher}}, \bibinfo {author} {\bibfnamefont {Kenneth}\
  \bibnamefont {Ratliff}}, \bibinfo {author} {\bibfnamefont {Scott}\
  \bibnamefont {Watson}}, \ and\ \bibinfo {author} {\bibfnamefont {Chris~D.}\
  \bibnamefont {White}},\ }\bibfield  {title} {\enquote {\bibinfo {title} {{The
  classical double copy in three spacetime dimensions}},}\ }\href {\doibase
  10.1007/JHEP07(2019)167} {\bibfield  {journal} {\bibinfo  {journal} {JHEP}\
  }\textbf {\bibinfo {volume} {07}},\ \bibinfo {pages} {167} (\bibinfo {year}
  {2019})},\ \Eprint {http://arxiv.org/abs/1904.11001} {arXiv:1904.11001
  [hep-th]} \BibitemShut {NoStop}%
\bibitem [{\citenamefont {Maybee}\ \emph {et~al.}(2019)\citenamefont {Maybee},
  \citenamefont {O'Connell},\ and\ \citenamefont {Vines}}]{Maybee:2019jus}%
  \BibitemOpen
  \bibfield  {author} {\bibinfo {author} {\bibfnamefont {Ben}\ \bibnamefont
  {Maybee}}, \bibinfo {author} {\bibfnamefont {Donal}\ \bibnamefont
  {O'Connell}}, \ and\ \bibinfo {author} {\bibfnamefont {Justin}\ \bibnamefont
  {Vines}},\ }\bibfield  {title} {\enquote {\bibinfo {title} {{Observables and
  amplitudes for spinning particles and black holes}},}\ }\href {\doibase
  10.1007/JHEP12(2019)156} {\bibfield  {journal} {\bibinfo  {journal} {JHEP}\
  }\textbf {\bibinfo {volume} {12}},\ \bibinfo {pages} {156} (\bibinfo {year}
  {2019})},\ \Eprint {http://arxiv.org/abs/1906.09260} {arXiv:1906.09260
  [hep-th]} \BibitemShut {NoStop}%
\bibitem [{\citenamefont {P.V.}\ and\ \citenamefont {Manu}(2020)}]{PV:2019uuv}%
  \BibitemOpen
  \bibfield  {author} {\bibinfo {author} {\bibfnamefont {Athira}\ \bibnamefont
  {P.V.}}\ and\ \bibinfo {author} {\bibfnamefont {A.}~\bibnamefont {Manu}},\
  }\bibfield  {title} {\enquote {\bibinfo {title} {{Classical double copy from
  Color Kinematics duality: A proof in the soft limit}},}\ }\href {\doibase
  10.1103/PhysRevD.101.046014} {\bibfield  {journal} {\bibinfo  {journal}
  {Phys. Rev. D}\ }\textbf {\bibinfo {volume} {101}},\ \bibinfo {pages}
  {046014} (\bibinfo {year} {2020})},\ \Eprint
  {http://arxiv.org/abs/1907.10021} {arXiv:1907.10021 [hep-th]} \BibitemShut
  {NoStop}%
\bibitem [{\citenamefont {Huang}\ \emph {et~al.}(2020)\citenamefont {Huang},
  \citenamefont {Kol},\ and\ \citenamefont {O'Connell}}]{Huang:2019cja}%
  \BibitemOpen
  \bibfield  {author} {\bibinfo {author} {\bibfnamefont {Yu-Tin}\ \bibnamefont
  {Huang}}, \bibinfo {author} {\bibfnamefont {Uri}\ \bibnamefont {Kol}}, \ and\
  \bibinfo {author} {\bibfnamefont {Donal}\ \bibnamefont {O'Connell}},\
  }\bibfield  {title} {\enquote {\bibinfo {title} {{Double copy of
  electric-magnetic duality}},}\ }\href {\doibase 10.1103/PhysRevD.102.046005}
  {\bibfield  {journal} {\bibinfo  {journal} {Phys. Rev. D}\ }\textbf {\bibinfo
  {volume} {102}},\ \bibinfo {pages} {046005} (\bibinfo {year} {2020})},\
  \Eprint {http://arxiv.org/abs/1911.06318} {arXiv:1911.06318 [hep-th]}
  \BibitemShut {NoStop}%
\bibitem [{\citenamefont {Alawadhi}\ \emph {et~al.}(2020)\citenamefont
  {Alawadhi}, \citenamefont {Berman}, \citenamefont {Spence},\ and\
  \citenamefont {Peinador~Veiga}}]{Alawadhi:2019urr}%
  \BibitemOpen
  \bibfield  {author} {\bibinfo {author} {\bibfnamefont {Rashid}\ \bibnamefont
  {Alawadhi}}, \bibinfo {author} {\bibfnamefont {David~S.}\ \bibnamefont
  {Berman}}, \bibinfo {author} {\bibfnamefont {Bill}\ \bibnamefont {Spence}}, \
  and\ \bibinfo {author} {\bibfnamefont {David}\ \bibnamefont
  {Peinador~Veiga}},\ }\bibfield  {title} {\enquote {\bibinfo {title}
  {{S-duality and the double copy}},}\ }\href {\doibase
  10.1007/JHEP03(2020)059} {\bibfield  {journal} {\bibinfo  {journal} {JHEP}\
  }\textbf {\bibinfo {volume} {03}},\ \bibinfo {pages} {059} (\bibinfo {year}
  {2020})},\ \Eprint {http://arxiv.org/abs/1911.06797} {arXiv:1911.06797
  [hep-th]} \BibitemShut {NoStop}%
\bibitem [{\citenamefont {Emond}\ \emph {et~al.}(2020)\citenamefont {Emond},
  \citenamefont {Huang}, \citenamefont {Kol}, \citenamefont {Moynihan},\ and\
  \citenamefont {O'Connell}}]{Emond:2020lwi}%
  \BibitemOpen
  \bibfield  {author} {\bibinfo {author} {\bibfnamefont {William~T.}\
  \bibnamefont {Emond}}, \bibinfo {author} {\bibfnamefont {Yu-tin}\
  \bibnamefont {Huang}}, \bibinfo {author} {\bibfnamefont {Uri}\ \bibnamefont
  {Kol}}, \bibinfo {author} {\bibfnamefont {Nathan}\ \bibnamefont {Moynihan}},
  \ and\ \bibinfo {author} {\bibfnamefont {Donal}\ \bibnamefont {O'Connell}},\
  }\bibfield  {title} {\enquote {\bibinfo {title} {{Amplitudes from Coulomb to
  Kerr-Taub-NUT}},}\ }\href@noop {} {\  (\bibinfo {year} {2020})},\ \Eprint
  {http://arxiv.org/abs/2010.07861} {arXiv:2010.07861 [hep-th]} \BibitemShut
  {NoStop}%
\bibitem [{\citenamefont {Berman}\ \emph {et~al.}(2020)\citenamefont {Berman},
  \citenamefont {Kim},\ and\ \citenamefont {Lee}}]{Berman:2020xvs}%
  \BibitemOpen
  \bibfield  {author} {\bibinfo {author} {\bibfnamefont {David~S.}\
  \bibnamefont {Berman}}, \bibinfo {author} {\bibfnamefont {Kwangeon}\
  \bibnamefont {Kim}}, \ and\ \bibinfo {author} {\bibfnamefont {Kanghoon}\
  \bibnamefont {Lee}},\ }\bibfield  {title} {\enquote {\bibinfo {title} {{The
  Classical Double Copy for M-theory from a Kerr-Schild Ansatz for Exceptional
  Field Theory}},}\ }\href@noop {} {\  (\bibinfo {year} {2020})},\ \Eprint
  {http://arxiv.org/abs/2010.08255} {arXiv:2010.08255 [hep-th]} \BibitemShut
  {NoStop}%
\bibitem [{\citenamefont {Naculich}(2014)}]{Naculich:2014naa}%
  \BibitemOpen
  \bibfield  {author} {\bibinfo {author} {\bibfnamefont {Stephen~G.}\
  \bibnamefont {Naculich}},\ }\bibfield  {title} {\enquote {\bibinfo {title}
  {{Scattering equations and BCJ relations for gauge and gravitational
  amplitudes with massive scalar particles}},}\ }\href {\doibase
  10.1007/JHEP09(2014)029} {\bibfield  {journal} {\bibinfo  {journal} {JHEP}\
  }\textbf {\bibinfo {volume} {09}},\ \bibinfo {pages} {029} (\bibinfo {year}
  {2014})},\ \Eprint {http://arxiv.org/abs/1407.7836} {arXiv:1407.7836
  [hep-th]} \BibitemShut {NoStop}%
\bibitem [{\citenamefont {Johansson}\ and\ \citenamefont
  {Ochirov}(2016)}]{Johansson:2015oia}%
  \BibitemOpen
  \bibfield  {author} {\bibinfo {author} {\bibfnamefont {Henrik}\ \bibnamefont
  {Johansson}}\ and\ \bibinfo {author} {\bibfnamefont {Alexander}\ \bibnamefont
  {Ochirov}},\ }\bibfield  {title} {\enquote {\bibinfo {title}
  {{Color-Kinematics Duality for QCD Amplitudes}},}\ }\href {\doibase
  10.1007/JHEP01(2016)170} {\bibfield  {journal} {\bibinfo  {journal} {JHEP}\
  }\textbf {\bibinfo {volume} {01}},\ \bibinfo {pages} {170} (\bibinfo {year}
  {2016})},\ \Eprint {http://arxiv.org/abs/1507.00332} {arXiv:1507.00332
  [hep-ph]} \BibitemShut {NoStop}%
\bibitem [{\citenamefont {Plefka}\ \emph {et~al.}(2020)\citenamefont {Plefka},
  \citenamefont {Shi},\ and\ \citenamefont {Wang}}]{Plefka:2019wyg}%
  \BibitemOpen
  \bibfield  {author} {\bibinfo {author} {\bibfnamefont {Jan}\ \bibnamefont
  {Plefka}}, \bibinfo {author} {\bibfnamefont {Canxin}\ \bibnamefont {Shi}}, \
  and\ \bibinfo {author} {\bibfnamefont {Tianheng}\ \bibnamefont {Wang}},\
  }\bibfield  {title} {\enquote {\bibinfo {title} {{Double copy of massive
  scalar QCD}},}\ }\href {\doibase 10.1103/PhysRevD.101.066004} {\bibfield
  {journal} {\bibinfo  {journal} {Phys. Rev. D}\ }\textbf {\bibinfo {volume}
  {101}},\ \bibinfo {pages} {066004} (\bibinfo {year} {2020})},\ \Eprint
  {http://arxiv.org/abs/1911.06785} {arXiv:1911.06785 [hep-th]} \BibitemShut
  {NoStop}%
\bibitem [{\citenamefont {Kosmopoulos}(2020)}]{Kosmopoulos:2020pcd}%
  \BibitemOpen
  \bibfield  {author} {\bibinfo {author} {\bibfnamefont {Dimitrios}\
  \bibnamefont {Kosmopoulos}},\ }\bibfield  {title} {\enquote {\bibinfo {title}
  {{Simplifying $D$-Dimensional Physical-State Sums in Gauge Theory and
  Gravity}},}\ }\href@noop {} {\  (\bibinfo {year} {2020})},\ \Eprint
  {http://arxiv.org/abs/2009.00141} {arXiv:2009.00141 [hep-th]} \BibitemShut
  {NoStop}%
\bibitem [{\citenamefont {Badger}\ \emph {et~al.}(2005)\citenamefont {Badger},
  \citenamefont {Glover}, \citenamefont {Khoze},\ and\ \citenamefont
  {Svrcek}}]{Badger:2005zh}%
  \BibitemOpen
  \bibfield  {author} {\bibinfo {author} {\bibfnamefont {S.~D.}\ \bibnamefont
  {Badger}}, \bibinfo {author} {\bibfnamefont {E.~W.~Nigel}\ \bibnamefont
  {Glover}}, \bibinfo {author} {\bibfnamefont {V.~V.}\ \bibnamefont {Khoze}}, \
  and\ \bibinfo {author} {\bibfnamefont {P.}~\bibnamefont {Svrcek}},\
  }\bibfield  {title} {\enquote {\bibinfo {title} {{Recursion relations for
  gauge theory amplitudes with massive particles}},}\ }\href {\doibase
  10.1088/1126-6708/2005/07/025} {\bibfield  {journal} {\bibinfo  {journal}
  {JHEP}\ }\textbf {\bibinfo {volume} {07}},\ \bibinfo {pages} {025} (\bibinfo
  {year} {2005})},\ \Eprint {http://arxiv.org/abs/hep-th/0504159}
  {arXiv:hep-th/0504159 [hep-th]} \BibitemShut {NoStop}%
\bibitem [{\citenamefont {Luna}\ \emph
  {et~al.}(2018{\natexlab{b}})\citenamefont {Luna}, \citenamefont {Nicholson},
  \citenamefont {O'Connell},\ and\ \citenamefont {White}}]{Luna2017dtq}%
  \BibitemOpen
  \bibfield  {author} {\bibinfo {author} {\bibfnamefont {Andr{\'e}s}\
  \bibnamefont {Luna}}, \bibinfo {author} {\bibfnamefont {Isobel}\ \bibnamefont
  {Nicholson}}, \bibinfo {author} {\bibfnamefont {Donal}\ \bibnamefont
  {O'Connell}}, \ and\ \bibinfo {author} {\bibfnamefont {Chris~D.}\
  \bibnamefont {White}},\ }\bibfield  {title} {\enquote {\bibinfo {title}
  {{Inelastic black hole scattering from charged scalar amplitudes}},}\ }\href
  {\doibase 10.1007/JHEP03(2018)044} {\bibfield  {journal} {\bibinfo  {journal}
  {JHEP}\ }\textbf {\bibinfo {volume} {03}},\ \bibinfo {pages} {044} (\bibinfo
  {year} {2018}{\natexlab{b}})},\ \Eprint {http://arxiv.org/abs/1711.03901}
  {arXiv:1711.03901 [hep-th]} \BibitemShut {NoStop}%
\bibitem [{\citenamefont {Carrasco}\ and\ \citenamefont
  {Johansson}(2011)}]{Carrasco:2011hw}%
  \BibitemOpen
  \bibfield  {author} {\bibinfo {author} {\bibfnamefont {John Joseph~M.}\
  \bibnamefont {Carrasco}}\ and\ \bibinfo {author} {\bibfnamefont {Henrik}\
  \bibnamefont {Johansson}},\ }\bibfield  {title} {\enquote {\bibinfo {title}
  {{Generic multiloop methods and application to N=4 super-Yang-Mills}},}\
  }\href {\doibase 10.1088/1751-8113/44/45/454004} {\bibfield  {journal}
  {\bibinfo  {journal} {J. Phys.}\ }\textbf {\bibinfo {volume} {A44}},\
  \bibinfo {pages} {454004} (\bibinfo {year} {2011})},\ \Eprint
  {http://arxiv.org/abs/1103.3298} {arXiv:1103.3298 [hep-th]} \BibitemShut
  {NoStop}%
\bibitem [{\citenamefont {Holstein}\ and\ \citenamefont
  {Ross}(2008)}]{Holstein:2008sx}%
  \BibitemOpen
  \bibfield  {author} {\bibinfo {author} {\bibfnamefont {Barry~R.}\
  \bibnamefont {Holstein}}\ and\ \bibinfo {author} {\bibfnamefont {Andreas}\
  \bibnamefont {Ross}},\ }\bibfield  {title} {\enquote {\bibinfo {title} {{Spin
  Effects in Long Range Gravitational Scattering}},}\ }\href@noop {} {\
  (\bibinfo {year} {2008})},\ \Eprint {http://arxiv.org/abs/0802.0716}
  {arXiv:0802.0716 [hep-ph]} \BibitemShut {NoStop}%
\bibitem [{\citenamefont {Vaidya}(2015)}]{Vaidya:2014kza}%
  \BibitemOpen
  \bibfield  {author} {\bibinfo {author} {\bibfnamefont {Varun}\ \bibnamefont
  {Vaidya}},\ }\bibfield  {title} {\enquote {\bibinfo {title} {{Gravitational
  spin Hamiltonians from the S matrix}},}\ }\href {\doibase
  10.1103/PhysRevD.91.024017} {\bibfield  {journal} {\bibinfo  {journal} {Phys.
  Rev. D}\ }\textbf {\bibinfo {volume} {91}},\ \bibinfo {pages} {024017}
  (\bibinfo {year} {2015})},\ \Eprint {http://arxiv.org/abs/1410.5348}
  {arXiv:1410.5348 [hep-th]} \BibitemShut {NoStop}%
\bibitem [{\citenamefont {Guevara}(2019)}]{Guevara:2017csg}%
  \BibitemOpen
  \bibfield  {author} {\bibinfo {author} {\bibfnamefont {Alfredo}\ \bibnamefont
  {Guevara}},\ }\bibfield  {title} {\enquote {\bibinfo {title} {{Holomorphic
  classical limit for spin effects in gravitational and electromagnetic
  scattering}},}\ }\href {\doibase 10.1007/JHEP04(2019)033} {\bibfield
  {journal} {\bibinfo  {journal} {JHEP}\ }\textbf {\bibinfo {volume} {04}},\
  \bibinfo {pages} {033} (\bibinfo {year} {2019})},\ \Eprint
  {http://arxiv.org/abs/1706.02314} {arXiv:1706.02314 [hep-th]} \BibitemShut
  {NoStop}%
\bibitem [{\citenamefont {Bini}\ and\ \citenamefont
  {Damour}(2017)}]{Bini:2017xzy}%
  \BibitemOpen
  \bibfield  {author} {\bibinfo {author} {\bibfnamefont {Donato}\ \bibnamefont
  {Bini}}\ and\ \bibinfo {author} {\bibfnamefont {Thibault}\ \bibnamefont
  {Damour}},\ }\bibfield  {title} {\enquote {\bibinfo {title} {{Gravitational
  spin-orbit coupling in binary systems, post-Minkowskian approximation and
  effective one-body theory}},}\ }\href {\doibase 10.1103/PhysRevD.96.104038}
  {\bibfield  {journal} {\bibinfo  {journal} {Phys. Rev. D}\ }\textbf {\bibinfo
  {volume} {96}},\ \bibinfo {pages} {104038} (\bibinfo {year} {2017})},\
  \Eprint {http://arxiv.org/abs/1709.00590} {arXiv:1709.00590 [gr-qc]}
  \BibitemShut {NoStop}%
\bibitem [{\citenamefont {Vines}\ \emph {et~al.}(2019)\citenamefont {Vines},
  \citenamefont {Steinhoff},\ and\ \citenamefont {Buonanno}}]{Vines:2018gqi}%
  \BibitemOpen
  \bibfield  {author} {\bibinfo {author} {\bibfnamefont {Justin}\ \bibnamefont
  {Vines}}, \bibinfo {author} {\bibfnamefont {Jan}\ \bibnamefont {Steinhoff}},
  \ and\ \bibinfo {author} {\bibfnamefont {Alessandra}\ \bibnamefont
  {Buonanno}},\ }\bibfield  {title} {\enquote {\bibinfo {title}
  {{Spinning-black-hole scattering and the test-black-hole limit at second
  post-Minkowskian order}},}\ }\href {\doibase 10.1103/PhysRevD.99.064054}
  {\bibfield  {journal} {\bibinfo  {journal} {Phys. Rev. D}\ }\textbf {\bibinfo
  {volume} {99}},\ \bibinfo {pages} {064054} (\bibinfo {year} {2019})},\
  \Eprint {http://arxiv.org/abs/1812.00956} {arXiv:1812.00956 [gr-qc]}
  \BibitemShut {NoStop}%
\bibitem [{\citenamefont {Guevara}\ \emph
  {et~al.}(2019{\natexlab{a}})\citenamefont {Guevara}, \citenamefont
  {Ochirov},\ and\ \citenamefont {Vines}}]{Guevara:2018wpp}%
  \BibitemOpen
  \bibfield  {author} {\bibinfo {author} {\bibfnamefont {Alfredo}\ \bibnamefont
  {Guevara}}, \bibinfo {author} {\bibfnamefont {Alexander}\ \bibnamefont
  {Ochirov}}, \ and\ \bibinfo {author} {\bibfnamefont {Justin}\ \bibnamefont
  {Vines}},\ }\bibfield  {title} {\enquote {\bibinfo {title} {{Scattering of
  Spinning Black Holes from Exponentiated Soft Factors}},}\ }\href {\doibase
  10.1007/JHEP09(2019)056} {\bibfield  {journal} {\bibinfo  {journal} {JHEP}\
  }\textbf {\bibinfo {volume} {09}},\ \bibinfo {pages} {056} (\bibinfo {year}
  {2019}{\natexlab{a}})},\ \Eprint {http://arxiv.org/abs/1812.06895}
  {arXiv:1812.06895 [hep-th]} \BibitemShut {NoStop}%
\bibitem [{\citenamefont {Chung}\ \emph {et~al.}(2019)\citenamefont {Chung},
  \citenamefont {Huang}, \citenamefont {Kim},\ and\ \citenamefont
  {Lee}}]{Chung:2018kqs}%
  \BibitemOpen
  \bibfield  {author} {\bibinfo {author} {\bibfnamefont {Ming-Zhi}\
  \bibnamefont {Chung}}, \bibinfo {author} {\bibfnamefont {Yu-Tin}\
  \bibnamefont {Huang}}, \bibinfo {author} {\bibfnamefont {Jung-Wook}\
  \bibnamefont {Kim}}, \ and\ \bibinfo {author} {\bibfnamefont {Sangmin}\
  \bibnamefont {Lee}},\ }\bibfield  {title} {\enquote {\bibinfo {title} {{The
  simplest massive S-matrix: from minimal coupling to Black Holes}},}\ }\href
  {\doibase 10.1007/JHEP04(2019)156} {\bibfield  {journal} {\bibinfo  {journal}
  {JHEP}\ }\textbf {\bibinfo {volume} {04}},\ \bibinfo {pages} {156} (\bibinfo
  {year} {2019})},\ \Eprint {http://arxiv.org/abs/1812.08752} {arXiv:1812.08752
  [hep-th]} \BibitemShut {NoStop}%
\bibitem [{\citenamefont {Guevara}\ \emph
  {et~al.}(2019{\natexlab{b}})\citenamefont {Guevara}, \citenamefont
  {Ochirov},\ and\ \citenamefont {Vines}}]{Guevara:2019fsj}%
  \BibitemOpen
  \bibfield  {author} {\bibinfo {author} {\bibfnamefont {Alfredo}\ \bibnamefont
  {Guevara}}, \bibinfo {author} {\bibfnamefont {Alexander}\ \bibnamefont
  {Ochirov}}, \ and\ \bibinfo {author} {\bibfnamefont {Justin}\ \bibnamefont
  {Vines}},\ }\bibfield  {title} {\enquote {\bibinfo {title} {{Black-hole
  scattering with general spin directions from minimal-coupling amplitudes}},}\
  }\href {\doibase 10.1103/PhysRevD.100.104024} {\bibfield  {journal} {\bibinfo
   {journal} {Phys. Rev. D}\ }\textbf {\bibinfo {volume} {100}},\ \bibinfo
  {pages} {104024} (\bibinfo {year} {2019}{\natexlab{b}})},\ \Eprint
  {http://arxiv.org/abs/1906.10071} {arXiv:1906.10071 [hep-th]} \BibitemShut
  {NoStop}%
\bibitem [{\citenamefont {Johansson}\ and\ \citenamefont
  {Ochirov}(2019)}]{Johansson:2019dnu}%
  \BibitemOpen
  \bibfield  {author} {\bibinfo {author} {\bibfnamefont {Henrik}\ \bibnamefont
  {Johansson}}\ and\ \bibinfo {author} {\bibfnamefont {Alexander}\ \bibnamefont
  {Ochirov}},\ }\bibfield  {title} {\enquote {\bibinfo {title} {{Double copy
  for massive quantum particles with spin}},}\ }\href {\doibase
  10.1007/JHEP09(2019)040} {\bibfield  {journal} {\bibinfo  {journal} {JHEP}\
  }\textbf {\bibinfo {volume} {09}},\ \bibinfo {pages} {040} (\bibinfo {year}
  {2019})},\ \Eprint {http://arxiv.org/abs/1906.12292} {arXiv:1906.12292
  [hep-th]} \BibitemShut {NoStop}%
\bibitem [{\citenamefont {Chung}\ \emph {et~al.}(2020)\citenamefont {Chung},
  \citenamefont {Huang},\ and\ \citenamefont {Kim}}]{Chung:2019duq}%
  \BibitemOpen
  \bibfield  {author} {\bibinfo {author} {\bibfnamefont {Ming-Zhi}\
  \bibnamefont {Chung}}, \bibinfo {author} {\bibfnamefont {Yu-Tin}\
  \bibnamefont {Huang}}, \ and\ \bibinfo {author} {\bibfnamefont {Jung-Wook}\
  \bibnamefont {Kim}},\ }\bibfield  {title} {\enquote {\bibinfo {title}
  {{Classical potential for general spinning bodies}},}\ }\href {\doibase
  10.1007/JHEP09(2020)074} {\bibfield  {journal} {\bibinfo  {journal} {JHEP}\
  }\textbf {\bibinfo {volume} {09}},\ \bibinfo {pages} {074} (\bibinfo {year}
  {2020})},\ \Eprint {http://arxiv.org/abs/1908.08463} {arXiv:1908.08463
  [hep-th]} \BibitemShut {NoStop}%
\bibitem [{\citenamefont {Damgaard}\ \emph {et~al.}(2019)\citenamefont
  {Damgaard}, \citenamefont {Haddad},\ and\ \citenamefont
  {Helset}}]{Damgaard:2019lfh}%
  \BibitemOpen
  \bibfield  {author} {\bibinfo {author} {\bibfnamefont {Poul~H.}\ \bibnamefont
  {Damgaard}}, \bibinfo {author} {\bibfnamefont {Kays}\ \bibnamefont {Haddad}},
  \ and\ \bibinfo {author} {\bibfnamefont {Andreas}\ \bibnamefont {Helset}},\
  }\bibfield  {title} {\enquote {\bibinfo {title} {{Heavy Black Hole Effective
  Theory}},}\ }\href {\doibase 10.1007/JHEP11(2019)070} {\bibfield  {journal}
  {\bibinfo  {journal} {JHEP}\ }\textbf {\bibinfo {volume} {11}},\ \bibinfo
  {pages} {070} (\bibinfo {year} {2019})},\ \Eprint
  {http://arxiv.org/abs/1908.10308} {arXiv:1908.10308 [hep-ph]} \BibitemShut
  {NoStop}%
\bibitem [{\citenamefont {Bautista}\ and\ \citenamefont
  {Guevara}(2019)}]{Bautista:2019evw}%
  \BibitemOpen
  \bibfield  {author} {\bibinfo {author} {\bibfnamefont {Yilber~Fabian}\
  \bibnamefont {Bautista}}\ and\ \bibinfo {author} {\bibfnamefont {Alfredo}\
  \bibnamefont {Guevara}},\ }\bibfield  {title} {\enquote {\bibinfo {title}
  {{On the Double Copy for Spinning Matter}},}\ }\href@noop {} {\  (\bibinfo
  {year} {2019})},\ \Eprint {http://arxiv.org/abs/1908.11349} {arXiv:1908.11349
  [hep-th]} \BibitemShut {NoStop}%
\bibitem [{\citenamefont {Aoude}\ \emph {et~al.}(2020)\citenamefont {Aoude},
  \citenamefont {Haddad},\ and\ \citenamefont {Helset}}]{Aoude:2020onz}%
  \BibitemOpen
  \bibfield  {author} {\bibinfo {author} {\bibfnamefont {Rafael}\ \bibnamefont
  {Aoude}}, \bibinfo {author} {\bibfnamefont {Kays}\ \bibnamefont {Haddad}}, \
  and\ \bibinfo {author} {\bibfnamefont {Andreas}\ \bibnamefont {Helset}},\
  }\bibfield  {title} {\enquote {\bibinfo {title} {{On-shell heavy particle
  effective theories}},}\ }\href {\doibase 10.1007/JHEP05(2020)051} {\bibfield
  {journal} {\bibinfo  {journal} {JHEP}\ }\textbf {\bibinfo {volume} {05}},\
  \bibinfo {pages} {051} (\bibinfo {year} {2020})},\ \Eprint
  {http://arxiv.org/abs/2001.09164} {arXiv:2001.09164 [hep-th]} \BibitemShut
  {NoStop}%
\bibitem [{\citenamefont {Bern}\ \emph {et~al.}(2020)\citenamefont {Bern},
  \citenamefont {Luna}, \citenamefont {Roiban}, \citenamefont {Shen},\ and\
  \citenamefont {Zeng}}]{Bern:2020buy}%
  \BibitemOpen
  \bibfield  {author} {\bibinfo {author} {\bibfnamefont {Zvi}\ \bibnamefont
  {Bern}}, \bibinfo {author} {\bibfnamefont {Andres}\ \bibnamefont {Luna}},
  \bibinfo {author} {\bibfnamefont {Radu}\ \bibnamefont {Roiban}}, \bibinfo
  {author} {\bibfnamefont {Chia-Hsien}\ \bibnamefont {Shen}}, \ and\ \bibinfo
  {author} {\bibfnamefont {Mao}\ \bibnamefont {Zeng}},\ }\bibfield  {title}
  {\enquote {\bibinfo {title} {{Spinning Black Hole Binary Dynamics, Scattering
  Amplitudes and Effective Field Theory}},}\ }\href@noop {} {\  (\bibinfo
  {year} {2020})},\ \Eprint {http://arxiv.org/abs/2005.03071} {arXiv:2005.03071
  [hep-th]} \BibitemShut {NoStop}%
\end{thebibliography}%

\end{document}